\documentclass[twocolumn, numberedappendix]{openjournal}

\usepackage[english]{babel}
\usepackage{amsmath}
\usepackage{bm}
\usepackage{xcolor}
\definecolor{linkcolor}{rgb}{0.0,0.3,0.5}
\usepackage{xspace}  
\usepackage{etoolbox}      
             
\usepackage{siunitx}
\usepackage{orcidlink}    
\usepackage{hyperref}
\usepackage[T1]{fontenc}
\usepackage{lmodern}
\hypersetup{
    unicode,
    colorlinks=true,
    linkcolor=linkcolor,
    citecolor=linkcolor,
    filecolor=linkcolor,
    urlcolor=linkcolor,
}
\makeatletter
  \patchcmd{\NAT@citex}
    {\@citea\NAT@hyper@{%
      \NAT@nmfmt{\NAT@nm}%
      \hyper@natlinkbreak{\NAT@aysep\NAT@spacechar}{\@citeb\@extra@b@citeb}%
      \NAT@date}}
    {\@citea\NAT@nmfmt{\NAT@nm}%
    \NAT@aysep\NAT@spacechar\NAT@hyper@{\NAT@date}}{}{}
  \patchcmd{\NAT@citex}
    {\@citea\NAT@hyper@{%
      \NAT@nmfmt{\NAT@nm}%
      \hyper@natlinkbreak{\NAT@spacechar\NAT@@open\if*#1*\else#1\NAT@spacechar\fi}%
        {\@citeb\@extra@b@citeb}%
      \NAT@date}}
    {\@citea\NAT@nmfmt{\NAT@nm}%
    \NAT@spacechar\NAT@@open\if*#1*\else#1\NAT@spacechar\fi\NAT@hyper@{\NAT@date}}
    {}{}
\makeatother

\newcommand\blfootnote[1]{%
  \begingroup
  \renewcommand\thefootnote{}\footnote{#1}%
  \addtocounter{footnote}{-1}%
  \endgroup
}
\providecommand{\uat}[2]{#1}
\providecommand{\gridline}[1]{\centering #1\par}
\providecommand{\facilities}[1]{\par\noindent\textit{Facilities:} #1\par}
\providecommand{\software}[1]{\par\noindent\textit{Software:} #1\par}
\newcommand{\reportnum}[1]{\blfootnote{#1}}
\DeclareSIUnit{\mag}{mag}
\DeclareSIUnit{\arcmin}{arcmin}
\DeclareSIUnit{\arcsec}{arcsec}
\DeclareSIUnit{\pc}{pc}
\DeclareSIUnit{\kpc}{kpc}
\DeclareSIUnit{\mas}{mas}
\DeclareSIUnit{\year}{yr}
\DeclareSIUnit{\dex}{dex}

\newcommand{\satah}{1.68^{+0.39}_{-0.30}\,\si{\arcmin}}
\newcommand{\satrh}{1.22^{+0.23}_{-0.19}\,\si{\arcmin}}

\newcommand{\satphysrh}{53^{+10}_{-8}\,\si{\pc}}
\newcommand{\satpa}{-68^{+12}_{-10}\,\si{\deg}}
\newcommand{\satell}{0.48^{+0.11}_{-0.13}}
\newcommand{\satmv}{-2.72^{+0.49}_{-0.70}\,\si{\mag}}

\newcommand{\satmod}{20.90^{+0.10}_{-0.11}}
\newcommand{\satheliodist}{151^{+8}_{-8}\,\si{\kpc}}
\newcommand{\satnstar}{114^{+11}_{-11}}

\newcommand{\satfeh}{-1.94\, \si{\dex}}

\begin{document}
\reportnum{FERMILAB-PUB-26-0487-PPD}
\title{\large A Deep Look at the Ultra-Faint Milky Way Satellite Virgo~III \\ with Rubin Observatory Data Preview 2}

\author{Aashay Pai$^{1,2,3,4,\star}$\orcidlink{0009-0008-9641-6065}}
\author{Alex Drlica-Wagner$^{2,3,5,6}$\orcidlink{0000-0001-8251-933X}}
\author{Peter S. Ferguson$^{7}$\orcidlink{0000-0001-6957-1627}}
\author{William Cerny$^{8}$\orcidlink{0000-0003-1697-7062}}
\author{\\Chin~Yi~Tan$^{1,2,3}$\orcidlink{0000-0003-0478-0473}}
\author{Jeffrey L. Carlin$^{9}$\orcidlink{0000-0002-3936-9628}}
\author{Yumi Choi$^{9}$\orcidlink{0000-0003-1680-1884}}
\author{\v{Z}eljko Ivezi\'{c}$^{9, 10}$\orcidlink{0000-0001-5250-2633}}
\author{\\Andr\'es~A.~Plazas~Malag\'on$^{11, 12}$\orcidlink{0000-0002-2598-0514}}
\author{Lauren~A.~MacArthur$^{13}$\orcidlink{0009-0003-5548-6773}}
\author{Colin~T.~Slater$^{10}$\orcidlink{0000-0002-0558-0521}}
\author{\\Dan~S.~Taranu$^{13}$\orcidlink{0000-0001-6268-1882}}
\author{Brian Yanny$^5$\orcidlink{0000-0002-9541-2678}}

\affiliation{$^1$Department of Physics, University of Chicago, Chicago, IL 60637, USA \\
$^2$Kavli Institute for Cosmological Physics, University of Chicago, Chicago, IL 60637, USA \\
$^3$NSF-Simons AI Institute for the Sky (SkAI), 172 E. Chestnut St., Chicago, IL 60611, USA \\
$^4$Universit\'e Grenoble Alpes, CNRS/IN2P3, LPSC, 53 avenue des Martyrs, F-38026 Grenoble, France \\
$^5$Fermi National Accelerator Laboratory, P.\ O.\ Box 500, Batavia, IL 60510, USA \\
$^6$Department of Astronomy and Astrophysics, University of Chicago, Chicago, IL 60637, USA \\
$^7$DIRAC Institute, Department of Astronomy, University of Washington, 3910 15th Avenue NE, Seattle, WA 98195, USA \\
$^8$Department of Astronomy, Yale University, New Haven, CT 06520, USA \\
$^9$NSF-DOE Vera C.\ Rubin Observatory / NSF NOIRLab, 950 N.\ Cherry Ave., Tucson, AZ 85719, USA \\
$^{10}$University of Washington, Dept.\ of Astronomy, Box 351580, Seattle, WA 98195, USA \\
$^{11}$SLAC National Accelerator Laboratory, 2575 Sand Hill Rd., Menlo Park, CA 94025, USA \\
$^{12}$Kavli Institute for Particle Astrophysics and Cosmology, SLAC National Accelerator Laboratory, 2575 Sand Hill Rd., Menlo Park, CA 94025, USA\\
$^{13}$Department of Astrophysical Sciences, Princeton University, Princeton, NJ 08544, USA}

\blfootnote{$^{\star}$\mbox{Email: \href{mailto:pai@uchicago.edu}{pai@uchicago.edu}}}

\begin{abstract}
We analyze the ultra-faint Milky Way satellite Virgo~III using data from the Vera C.\ Rubin Observatory Data Preview 2 (DP2).
Virgo~III was observed in the Rubin `Cosmic Treasure Chest' (M49) First Look field, which contains 924 visits in the {\it ugri} bands comprising $\sim \SI{10.5}{hrs}$ of exposure time with LSSTCam.
These data are considerably deeper than the majority of DP2, with a $5\sigma$ limiting magnitude that approaches the expected 10-year depth of LSST ($\sim25.2$--$26.5$\,mag, depending on band).
We report the morphological and stellar population parameters of Virgo~III measured with the maximum-likelihood-based package \texttt{ugali}.
The depth of the Rubin imaging yields more than a factor of four increase in the number of candidate member stars ($N_* = \satnstar$) relative to the Virgo~III discovery results ($N_* = 25^{+5}_{-4}$), enabling significantly more precise morphological constraints.
Our best-fit parameters broadly agree with previous measurements, further confirming that Virgo~III has properties that are consistent with an ultra-faint dwarf galaxy ($M_V = \satmv$; $r_{1/2} = \satphysrh$) located at a heliocentric distance of $D_\odot = \satheliodist$.
We also demonstrate that the depth and photometric quality of the DP2 data are sufficient to separate metal-poor and metal-rich stars in color--color space.
We further present period estimates for the three known RR Lyrae member stars derived from the DP2 forced photometry, and we use theoretical Period-Luminosity-Metallicity (PLZ) and Period-Wesenheit-Metallicity (PWZ) relations to obtain independent distance estimates.
We find that our period and distance estimates are broadly consistent with previous measurements for these RR Lyrae.
These results demonstrate the power of LSST data for the discovery and characterization of ultra-faint dwarf galaxies and motivate future searches for new satellites across the southern sky.
\end{abstract}


\section{Introduction} 
\label{sec:intro} 
The NSF-DOE Vera C.\ Rubin Observatory, funded by the U.S.\ National Science Foundation and the U.S.\ Department of Energy's Office of Science, is located on Cerro Pach\'{o}n in Chile. 
It began the ten-year Legacy Survey of Space and Time (LSST) in June 2026, during which the LSST Camera \citep[LSSTCam;][]{slac2025} will image the sky visible from Chile $\sim$800 times, resulting in measurements of $\sim$20 billion galaxies and $\sim$10 billion stars \citep[][]{ivezic2019}. 
The ``wide-fast-deep'' (WFD) survey strategy aims to observe the entire accessible southern sky once every three days, generating data with an unprecedented level of photometric precision and temporal sampling in six bands ({\it ugrizy}).
The LSST data are expected to transform many fields of astronomy, including the domain of near-field cosmology. In particular, LSST will enable the discovery of ultra-faint dwarf galaxies (UFDs) that are fainter and more distant than most of the known UFD population \citep[e.g.,][]{tollerud2008, hargis2014, tsiane2025}.

UFDs are the oldest ($\sim$\SI{12}{Gyr}), faintest ($M_V \gtrsim -7.7$; $L_\star \lesssim 10^5\,L_\odot$), and most metal-poor ([Fe/H] $\lesssim -2$) galaxies \citep{simon2019}.
They are extremely dark-matter-dominated systems, making them excellent probes of the properties of dark matter \citep[e.g.,][]{2010MNRAS.406.1220W}.
Recent sky surveys such as Pan-STARRS1 \citep[PS1;][]{chambers2016}, the Dark Energy Survey \citep[DES;][]{abbott2016}, the DECam Local Volume Exploration survey \citep[DELVE;][]{drlicawagner2021}, and the Hyper Suprime-Cam Subaru Strategic Program \citep[HSC-SSP;][]{Aihara:2018} have significantly increased the known population of UFDs extending beyond the Milky Way virial radius ($\sim$\SI{300}{kpc}).
Furthermore, these contiguous, wide-area programs have enabled rigorous determinations of survey selection functions \citep[e.g.,][]{2008ApJ...686..279K, 2009AJ....137..450W, drlicawagner2020, tan:2026}, which in turn have led to statistical analyses of the UFD population and detailed modeling of dark matter substructure and galaxy formation \citep[e.g.,][]{2018MNRAS.473.2060J, 2020ApJ...893...48N, 2022MNRAS.515.3685S, 2022MNRAS.516.3944M}.

Several studies have made projections for the Milky Way satellite galaxy population that will be observable with LSST.
\citet{tollerud2008} predicted that LSST will observe between $145$--$283$ systems as faint as $\sim 100\,L_\odot$ out to distances of $\sim$\SI{200}{\kpc}, and that the coadded images will be deep enough to detect all Milky Way satellite galaxies brighter than Bo\"otes~II ($M_V \sim -2.7$; $r_{1/2} \sim \SI{50}{\pc}$).
\citet{hargis2014} subsequently predicted an observable population of $37$--$114$ classical dwarfs ($L > 10^3\,L_\odot$) and $131$--$782$ hyperfaint dwarfs ($L < 10^3\,L_\odot$) out to $\sim$\SI{300}{\kpc}.
More recent estimates using updated luminosity functions predicted that LSST will detect $\sim$60 systems with $M_V < 0$ \citep{newton2018} and $\sim$80 systems with $-1 > M_V > -11$ \citep{jethwa2018}.
\citet{2022MNRAS.516.3944M} predicted that the LSST Y1 survey will detect all $273^{+119}_{-92}$ Milky Way satellites with $M_V < 0$, $r_{1/2} > \SI{10}{\pc}$, and $\mu_V < \SI{32}{\mag \per \arcsec^2}$ out to a heliocentric distance of $\SI{300}{\kpc}$.
Recently, \citet{tsiane2025} predicted that the WFD component of LSST will enable the detection of $\sim$90 satellites with $M_V < 0$ and $r_{1/2} > \SI{10}{\pc}$ out to an assumed virial radius of $\sim$\SI{300}{\kpc}.
A key takeaway from \citet{tsiane2025} is that reliable star--galaxy separation is essential to achieving the predicted detection efficiency, particularly at faint magnitudes.
Finally, \citet{rusterucci2026} investigated the impact of combining information from more than two photometric bands, and predicted that LSST would discover $\sim 50$ new Milky Way satellites, comparable to the estimates by \citet{tsiane2025}.

Here, we explore the sensitivity of the as-built Rubin system using observations collected during commissioning and early operations. These observations fortuitously include the Milky Way satellite Virgo~III, a UFD candidate discovered in data from HSC-SSP located at (R.A., Decl.) = (\ang{186.3}, \ang{4.4}) \citep{homma2024}.
It has an estimated $V$-band absolute magnitude of $M_V \sim -2.69$, a heliocentric distance of $D_\odot \sim \SI{151}{kpc}$, and a half-light radius of $r_{1/2} \sim \SI{44}{pc}$.
\citet{ngeow2024} also discovered three RR Lyrae stars (two type AB and one type C) that are consistent with the distance estimate above.
We are not aware of astrometric or spectroscopic studies of Virgo~III in the literature.

Virgo~III resides within the Rubin First Look ``Cosmic Treasure Chest'' field centered on the nearby Virgo galaxy cluster (M49). 
This field has been processed and included in Rubin DP2, thus making Virgo~III an ideal benchmark for developing and validating photometric analysis tools for the broader study of UFDs using deep imaging from LSST, including morphological characterization, photometric metallicity estimates, and distance measurements from standard candles such as RR Lyrae stars.
In this paper, we present the adapted capabilities of one such tool, \texttt{ugali}\footnote{\url{https://github.com/DarkEnergySurvey/ugali}}, which has been used to characterize the morphology of many faint satellite systems \citep[e.g.,][]{bechtol2015, drlicawagner2020, cerny2023, tan2025b}.

This paper is structured as follows.
Section~\ref{sec:data} describes the Rubin observations of the M49 field, the DP2 data products, and a high-level summary of the data quality.
Section~\ref{sec:morph} reports the properties of Virgo~III measured with the DP2 data and a comparison to previous measurements.
Section~\ref{sec:mem_properties} explores properties of the Virgo~III member stars, including their proper motions, their distribution in color--color space, and their photometric metallicities.
Section~\ref{sec:rrl} presents the analysis of the DP2 light curves of the known RR Lyrae stars.
The implications of our study of Virgo~III are discussed in Section~\ref{sec:discussion}.

\section{Rubin Data}
\label{sec:data} 
The Rubin DP2 \citep{dp2} contains 16,790 science-grade exposures, which have been processed to produce coadd images and object catalogs.
This data set covers up to $\sim$\SI{3000}{\deg^2} depending on band.
Here, we summarize the LSSTCam observations of the M49 field and briefly outline the LSST data processing pipelines. We also describe the DP2 data products that we use for the analysis in subsequent sections.

\subsection{Observations}
\label{sec:obs} 
\begin{figure*}[t!]
    \centering
\includegraphics[width=0.95\textwidth]{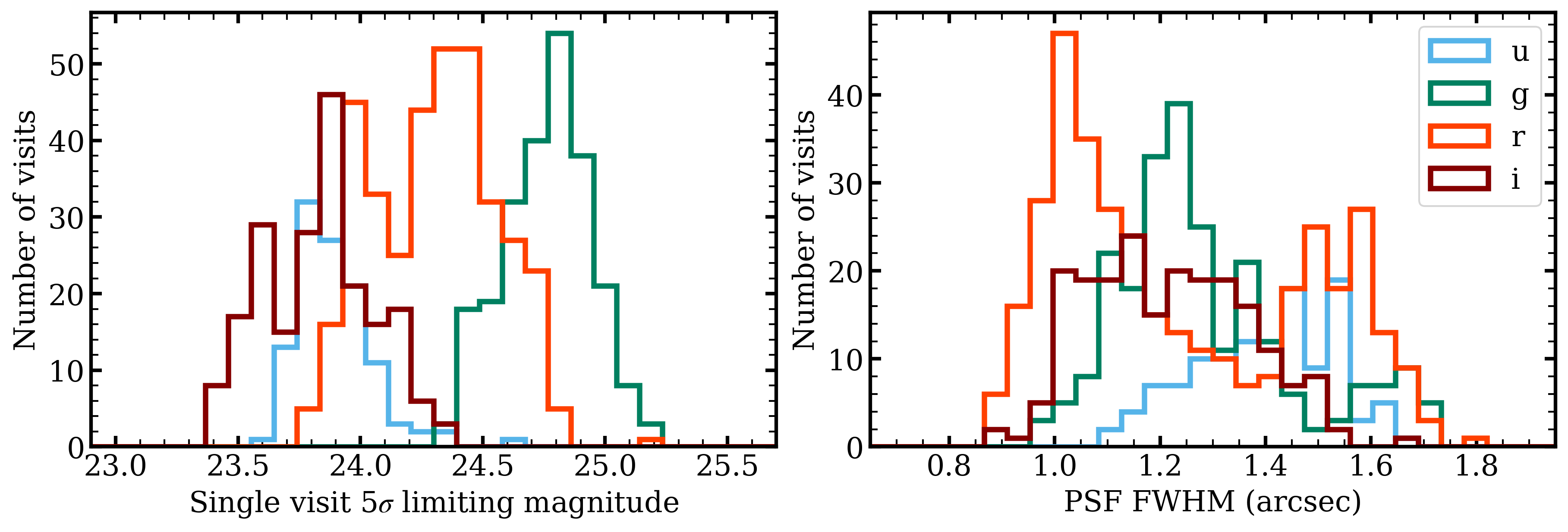}
    \caption{Distributions of per-visit $5\sigma$ limiting magnitude and PSF FWHM for the M49 field in Rubin DP2.}
    \label{fig:visit_summary_stats}
\end{figure*}

\begin{deluxetable*}{cccccc}
\tablecaption{\label{tab:visit_summary_stats} Summary statistics of the DP2 data for a \ang{2} radius around Virgo~III and for the entire M49 field. 
DP2 does not include any $z$ or $y$ band data for these fields.}

\tablehead{\colhead{Survey Property} & 
\colhead{Field} & \colhead{$u$} & \colhead{$g$} & \colhead{$r$} & \colhead{$i$}}
\startdata
Number of coadded visits & Virgo~III (\ang{2} radius) & 47 & 84 & 130 & 74 \\
Coadded median $5\sigma$ limiting magnitude & Virgo~III (\ang{2} radius) & 25.34 & 26.69 & 26.47 & 25.85 \\
\noalign{\vskip 0.6ex}\hline\noalign{\vskip 0.6ex}
Number of coadded visits & M49 & 119 & 236 & 361 & 208 \\
Median $5\sigma$ limiting magnitude per visit & M49 & 23.88 & 24.77 & 24.33 & 23.85 \\
Coadded median $5\sigma$ limiting magnitude & M49 & 25.27 & 26.49 & 26.02 & 25.67 \\
Median PSF FWHM (arcsec) & M49 & 1.42 & 1.25 & 1.17 & 1.21 \\
\hline
\enddata
\end{deluxetable*}
The Rubin DP2 M49 field (containing Virgo~III)  covers $\sim$\SI{24}{\deg^2} and is centered at (R.A., Decl.) = (\ang{186}, \ang{6.5}). 
It is one of 20 target fields that was observed during the LSSTCam commissioning campaign. 
Data for this field were collected on seven nights between 21 April 2025 and 3 May 2025 in support of the Rubin First Look event.

The LSSTCam observations for M49 include a total of 1106 visits: $N_u = 236$, $N_g = 278$, $N_r = 378$ and $N_i = 214$.
To be included in the coadded images, the single-exposure visit images need to pass several quality criteria.
These include a visit-level median point spread function (PSF) full-width at half maximum (FWHM) better than 1.7\,arcsec, a per-detector PSF FWHM better than 1.8\,arcsec and a transparency of at least 0.75.
Data from the eight corner detectors is also excluded due to heavy vignetting.
Finally, only visits from ``science'' programs were included in the coadds.
Out of the 1106 visits, 924 visits pass the quality criteria above: $N_u = 119$, $N_g = 236$, $N_r = 361$ and $N_i = 208$.
Visits in $u$-band had an exposure time of \SI{38}{\s}, while the exposure time for the $gri$ band visits was \SI{30}{\s}.
Additional details about the data selection for DP2 can be found in \citet{RTN-111} and \citet{RTN-115}.

Figure~\ref{fig:visit_summary_stats} shows the distribution of median $5\sigma$ limiting magnitudes and the PSF FWHM of the data pertaining to the M49 field.
Table~\ref{tab:visit_summary_stats} also shows a summary of these survey statistics.
The single-visit statistics (restricted to visits that passed the coaddition quality criteria) are taken from the LSSTCam Consolidated Database \citep[ConsDB;][]{DMTN-227}.
We calculate the coadded median $5\sigma$ limiting magnitude empirically using the \texttt{Object} catalogs (see next section for details) by selecting point-like sources ($\texttt{refExtendedness} = 0$; see Section~\ref{sec:sg_separation} for details) with a signal-to-noise ratio (SNR) between $4.75 < {\rm SNR} < 5.25$.
The coadded median $5\sigma$ magnitude for the M49 region is shallowest in the $u$-band and deepest in the $g$-band (Table~\ref{tab:visit_summary_stats}).

\subsection{\label{sec:dataproducts} DP2 Data Products}
The DP2 data were processed at the U.S. Data Facility (USDF) using the LSST Science Pipelines \citep{bosch2019, PSTN-019}. 
An overview of the DP2 data processing is provided in \citet{RTN-111} and \citet{RTN-115}.
The processing pipeline consists of four main steps.

In Step 1, instrument signature removal (ISR) is applied, including bias and dark subtraction, flat-fielding, cross-talk correction, and the removal of other instrumental artifacts. 
Source detection is then performed on the ISR-corrected images, producing initial photometric calibration through cross-matching to an external photometric reference catalog \citep{DMTN-277} and an initial PSF model using \texttt{PSFex} \citep{bertin2011}.

Step 2 performs an improved calibration of the data to ensure uniformity across the full dataset. 
Astrometric calibration is performed with \texttt{gbdes} \citep{bernstein2022}, photometric calibration with \texttt{FGCM} \citep{burke2018}, and PSF modeling with \texttt{PiFF} \citep{jarvis2021}.

Step 3 performs coaddition and measurement of single-visit images. 
The \texttt{deep\_coadd} images are produced by warping and stacking single-visit images with inverse-variance weighting, and sources extracted from these coadded images are stored in the \texttt{Object} catalogs.

Step 4 produces difference images and forced photometry, yielding the \texttt{DiaSource}, \texttt{DiaObject}, \texttt{ForcedSource} and \texttt{ForcedSourceOnDiaObject} tables. 

Unless otherwise specified, the analysis presented in 
Section~\ref{sec:morph} and Section~\ref{sec:mem_properties} is based on colors and magnitudes obtained from \texttt{psfFlux} measurements on the coadded images from the \texttt{Object} catalogs, and all magnitudes quoted are in the AB system \citep{oke1983}. 
To calculate the extinction due to dust,  we use $E(B-V)$ values obtained from the \citet{schlegel1998} maps, which are provided in the \texttt{Object} tables.
Following \citet{schlafly2011}, we correct these $E(B-V)$ values with a multiplicative factor of $0.86$.
We use $R_b$ values for the LSST bandpasses, which are calculated using the extinction curve from \citet{cardelli1989} with a flat spectral energy distribution assuming $R_V = 3.1$. These values are $R_u = 4.757$, $R_g = 3.661$, $R_r = 2.701$, $R_i = 2.054$, $R_z = 1.590$ and $R_y = 1.308$ \citep{lsst_tutorials}.\footnote{\url{https://dp1.lsst.io/tutorials/notebook/309/notebook-309-1.html}} 
The \texttt{psfFlux} measurements from the \texttt{ForcedSourceOnDiaObject} tables are used to construct light curves for the RR Lyrae analysis presented in Section~\ref{sec:rrl}.

\subsection{Star--Galaxy Separation}
\label{sec:sg_separation}
Stars and galaxies can be morphologically separated reliably when galaxies are sufficiently bright and/or spatially extended.
The main challenges to accurate star--galaxy separation arise at fainter magnitudes. In this regime, the galaxy population appears more compact and point-like, while lower signal-to-noise reduces the precision of size and shape measurements \citep{fadely2012, slater2020}. Furthermore, galaxies significantly outnumber stars at magnitudes fainter than $g \sim 21\,\si{mag}$.

Misclassification of galaxies as stars reduces the purity of the stellar sample and can bias inferences.
Contamination of the stellar sample is most severe at faint magnitudes, where the number density of unresolved, point-like galaxies rises steeply with depth.
A high-completeness and high-purity stellar sample is important for  UFD detection \citep[e.g.,][]{tsiane2025}.

Rubin DP2 provides several derived quantities that can be used for star--galaxy separation.
\begin{enumerate}
    \item \texttt{\{band\}\_extendedness} (binary) --- A binary parameter per band, defined in terms of the ratio of the PSF and composite model (cModel) flux measurements \citep{abazajian2004, bosch2018}.
    An object is classified as a point source (\texttt{\{band\}\_extendedness} $= 0$) if ${\rm flux}_{\rm psf} > 0.985 \times {\rm flux}_{\rm cmodel}$, and as extended (\texttt{\{band\}\_extendedness} $= 1$) otherwise.
    
    \item \texttt{refExtendedness} (binary) --- In the coaddition process, a reference band is chosen for each object based on its detection significance, in a fixed priority order $(i, r, z, y, g, u)$, where $i$ has the highest priority and $u$ has the lowest.
    This parameter is simply the \texttt{\{band\}\_extendedness} of the object in the reference band \citep{bosch2018}.
    
    \item \texttt{\{band\}\_sizeExtendedness} (floating point) --- This parameter is a quasi-probabilistic, likelihood-based comparison of the HSM moments \citep{hirata2003, mandelbaum2005} and PSF size of the object.
    
    \item \texttt{\{band\}\_model\_extendedness} (floating point) --- This parameter provides a S\'ersic flux and size-based estimate of whether an object is point-like or extended.
    
    \item \texttt{griz\_model\_extendedness} (floating point) --- This parameter provides a multiband ($griz$) S\'ersic model flux and size-based estimate of whether an object is a point source or extended.
\end{enumerate}
Further details about the classifiers are provided in the DP2 schema.\footnote{\url{https://sdm-schemas.lsst.io/dp2.html}}

\begin{figure}[t!]
    \centering
    \includegraphics[width=0.95\columnwidth]{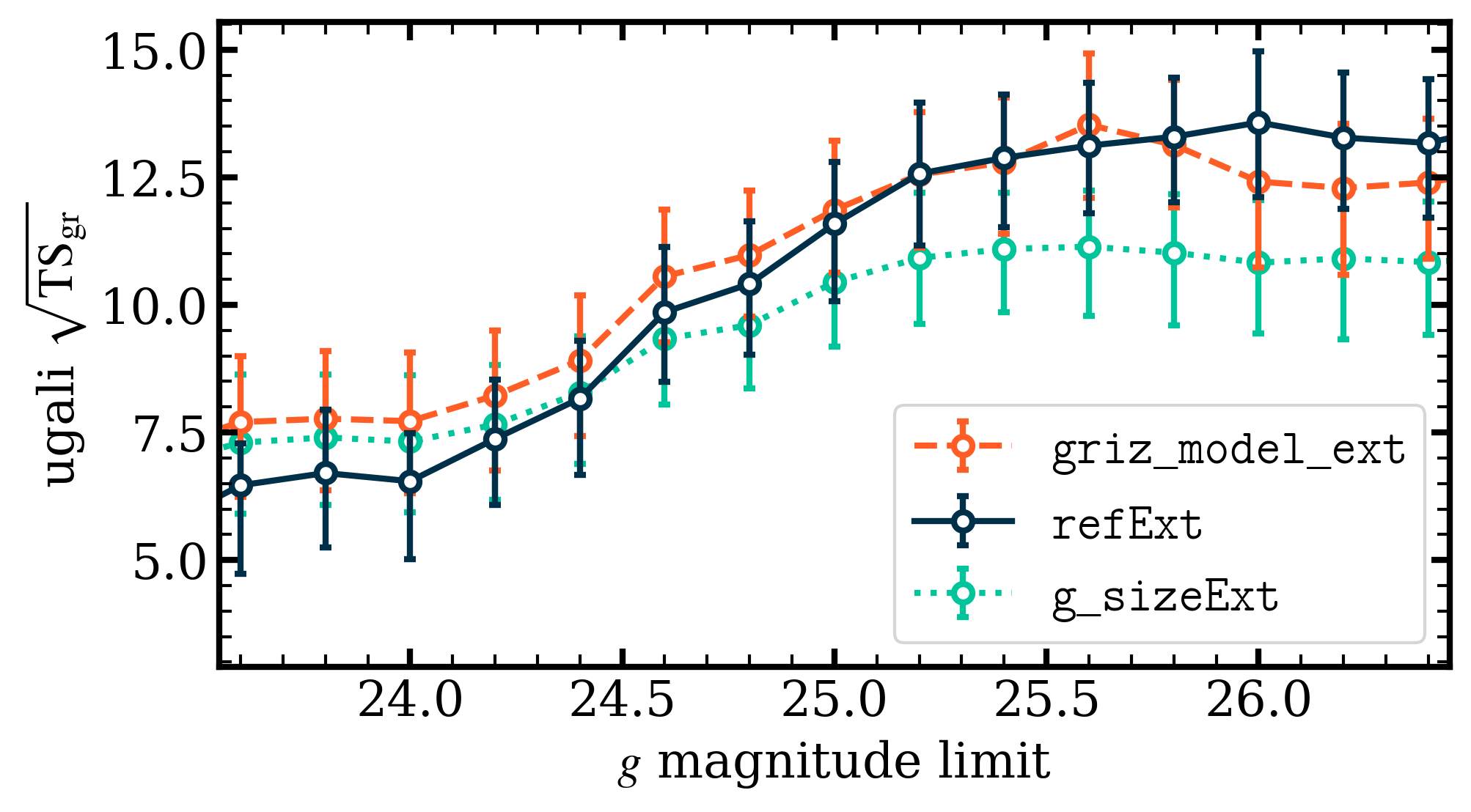}
    \caption{The \texttt{ugali} significance of the Virgo~III stellar overdensity  as a function of magnitude limit in the $g$-band for three different star--galaxy classifiers provided in the DP2 \texttt{Object} catalogs: \texttt{refExtendedness} (dark blue solid line), \texttt{g\_sizeExtendedness} (cyan dotted line) and \texttt{griz\_model\_extendedness} (orange dashed line).
    The \texttt{refExtendedness} classifier maximizes the ugali significance at $\sqrt{{\rm TS}_{gr}} \sim 14$ and performs the best at faint magnitudes.
    Uncertainties are estimated by bootstrapping sampling of the stellar catalog 500 times.}
    \label{fig:sg_classifier}
\end{figure}

We use the likelihood-based analysis package \texttt{ugali} to assess the performance of each star--galaxy classifier. 
Briefly, \texttt{ugali} performs a likelihood comparison between the null hypothesis containing only an empirically measured background (containing stars and misclassified galaxies) and an alternative hypothesis which also includes a resolved satellite template (See Section~\ref{sec:morph} for details).
We choose to use the star--galaxy classifier that maximizes the \texttt{ugali} detection significance for Virgo~III, which is defined in terms of the log-likelihood ratio (in $g$ and $r$ bands) ${\rm TS}_{gr} \equiv 2\,\Delta(\ln{\mathcal{L}})$.

We use morphological parameters from \citet{homma2024} and calculate the \texttt{ugali} significance ($\sqrt{ {\rm TS}_{gr}}$) of Virgo~III restricting the catalog to a range of faint-end $g$-band magnitude limits between $23$--$27\,\si{mag}$ in steps of \SI{0.2}{mag}. 
We repeat this procedure for multiple star--galaxy classifiers (\texttt{refExtendedness}, \texttt{g\_sizeExtendedness}, \texttt{griz\_model\_extendedness}, and \texttt{\{band\}\_model\_extendedness}), where the stars are selected as $\texttt{\{classifier\}} < 0.5$.
We do not use the \texttt{\{band\}\_extendedness} as it is incorporated into the \texttt{refExtendedness} shown here.

Figure~\ref{fig:sg_classifier} shows the resulting \texttt{ugali} significance as a function of magnitude limit for three classifiers.
The \texttt{ugali} significance of the individual \texttt{\{band\}\_model\_extendedness} follows the same trend as the \texttt{griz\_model\_extendedness} and is therefore not shown here.
We observe that the significance is maximized at $\sqrt{{\rm TS}_{gr}}\sim 14$ for both \texttt{refExtendedness} and \texttt{griz\_model\_extendedness}, but the former performs better at fainter magnitudes.\footnote{Note that the ${\rm TS}_{gr}$ can be converted to an asymptotic statistical significance using Wilks' theorem.}
Since we are interested in exploring the performance of LSST data at faint magnitudes, we opt to use \texttt{refExtendedness} as the baseline star--galaxy separator for subsequent analyses in this paper.

\begin{deluxetable*}{cccc}[t!]
\tablecaption{\label{tab:ugali_params} Fitted and derived parameters of Virgo III from \texttt{ugali}.}
\tablehead{
\colhead{Parameter} & 
\colhead{Description} & 
\colhead{Value} & 
\colhead{Units}
}
\startdata
\textbf{Morphological Parameters} & & & \\
$\alpha_{J2000}$ & Centroid right ascension & $186.350^{+0.002}_{-0.002}$ & deg \\
$\delta_{J2000}$ & Centroid declination & $4.442^{+0.003}_{-0.003}$ & deg \\
$a_h$ & Angular semi-major axis length & $1.68^{+0.39}_{-0.30}$ & arcmin \\
$r_h$ & Angular half-light radius (azimuthal avg.) & $1.22^{+0.23}_{-0.19}$ & arcmin \\
$\varepsilon$ & Ellipticity & $0.48^{+0.11}_{-0.13}$ & \nodata \\
PA & Major axis position angle & $-68^{+12}_{-10}$ & deg \\
$\lambda$ & Stellar richness & $5073^{+902}_{-849}$ & \nodata \\
\noalign{\vskip 0.6ex}\hline\noalign{\vskip 0.6ex}
\textbf{Isochrone Parameters} & & & \\
$(m-M)_0$ & Distance modulus$^a$ & $20.90^{+0.10}_{-0.11}$ & mag \\
$\tau$ & Isochrone age$^b$ & $> 11.45$ & Gyr \\
$Z$ & Isochrone metallicity$^b$ & $< 0.00018$  & \nodata \\
\noalign{\vskip 0.6ex}\hline\noalign{\vskip 0.6ex}
\textbf{Derived Parameters} & & & \\
$a_{1/2}$ & Physical semi-major axis length & $73^{+17}_{-13}$ & pc \\
$r_{1/2}$ & Physical half-light radius (azimuthal avg.) & $53^{+10}_{-8}$ & pc \\
$D_\odot$ & Heliocentric distance & $151^{+8}_{-8}$ & kpc \\
$[\mathrm{Fe/H}]$ & Isochrone metallicity & $< -1.94$ & dex \\
$M_V$ & Absolute $V$-band magnitude & $-2.72^{+0.49}_{-0.70}$ & mag \\
$L_V$ & $V$-band luminosity & $1654^{+443}_{-463}$ & $L_\odot$ \\
$M_*$ & Stellar mass (with $M_*/L_V = 2$) & $3307^{+886}_{-925}$ & $M_\odot$ \\
$N_*$ & Number of detected stars & $114^{+11}_{-11}$ & \nodata \\
\hline
\enddata
\tablenotetext{a}{\centering This includes a systematic of \SI{0.1}{mag} to account for the uncertainty arising from the isochrone modelling.}
\tablenotetext{b}{\centering The age and metallicity posteriors were distributed against the allowed MCMC bounds. 
We therefore quote the 84\% credible lower/upper limits.}
\end{deluxetable*}

\section{Morphological Characterization}
\label{sec:morph} 
\subsection{Maximum-Likelihood Fit}
\label{sec:morph_fit} 

We use the maximum-likelihood-based Ultra-faint GAlaxy LIkelihood (\texttt{ugali}) package \citep{bechtol2015, drlicawagner2020} to estimate the morphological parameters of Virgo~III.

We use an elliptical \citet{plummer1911} profile to fit the stellar density profile of Virgo~III.
The spatial model has the following parameters: the centroid coordinates $(\alpha_{J2000}, \delta_{J2000})$, the angular semi-major axis of the elliptical half-light radius $a_h$, the ellipticity $\varepsilon$, the position angle (PA) of the major axis, and the stellar richness $\lambda$.
We use \texttt{PARSEC-COLIBRI} isochrones \citep{bressan2012, tang2014, chen2014, chen2015, marigo2017} generated using the LSST bandpasses\footnote{LSST throughputs Release 1.9 (Sept.\ 2023): \url{https://stev.oapd.inaf.it/cmd_3.9/photsys.html}} to fit the measured magnitudes and colors of the stars.
We also use the \citet{chabrier2001} initial mass function.
The isochrone model has the following free parameters: the distance modulus $\mu = (m-M)_0$, the age $\tau$, and the metallicity $Z$.
We only use stars with $g$ and $r$-band magnitudes brighter than \SI{26}{mag} due to the lack of completeness and the large uncertainties at fainter magnitudes.
The nine free parameters are fit simultaneously using the Markov Chain Monte Carlo sampler \texttt{emcee} \citep{foremanmackey2013}.
The posterior distributions are shown in Appendix~\ref{app:ugali_mcmc}.

\begin{figure*}[t!]
    \centering
    \includegraphics[width=0.95\textwidth]{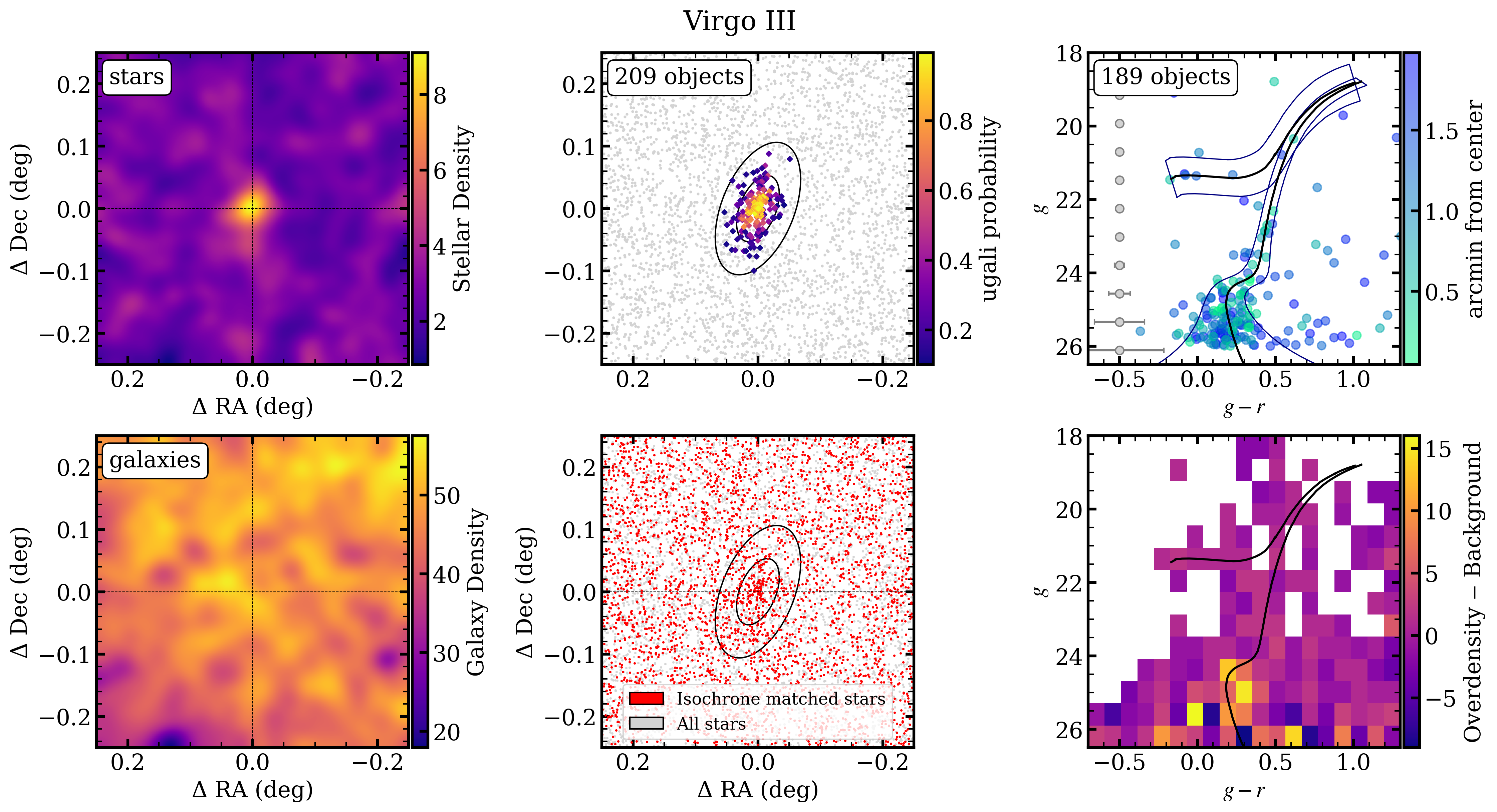}
    \caption{Diagnostic plots for Virgo~III observed in Rubin DP2.
    The top-left panel shows a smoothed map of isochrone-selected stellar density, with the overdensity in the center corresponding to Virgo~III.
    The top-middle panel shows the spatial distribution of isochrone-selected stars (gray) along with probable members colored by their \texttt{ugali} membership probability.
    The two black ellipses correspond to $2a_h$ and $4a_h$.
    The top-right panel shows $g$ vs.\ $g-r$ for stars within $\SI{2}{\arcmin}$ of Virgo~III, colored by their angular distance from the fitted centroid; the isochrone filter selects 189 out of the 209 stars with \texttt{ugali} membership probability $>10\%$.
    The bottom-left panel shows a smoothed map of isochrone-selected galaxy density to look for anomalies in the catalog data.
    The bottom-middle panel shows the spatial distribution of all stars in gray and isochrone-filtered stars in red.
    The bottom-right panel shows a Hess diagram of stars within $\SI{4}{\arcmin}$ of Virgo~III relative to a background annulus from $20$\,arcmin to $\SI{20.4}{\arcmin}$.}
    \label{fig:gr_6panel}
\end{figure*}

The \texttt{ugali} significance of the stellar overdensity corresponding to Virgo~III is $\sqrt{{\rm TS}_{gr}} = 15$ (Note that this value is different from the previous section as this uses the best fit parameters from the MCMC).
Table~\ref{tab:ugali_params} shows the measured values of the stellar density and isochrone parameters from \texttt{ugali}.
The parameter estimates are obtained from the median of the marginalized posteriors, while the uncertainties represent the 16th and 84th percentile values of the posterior distributions.
The table also includes derived morphological quantities.
The azimuthally averaged angular half-light radius, $r_{h}$, is defined as
\begin{equation}
\label{eq:rh}
    r_{h} = a_{h}\sqrt{1-\varepsilon}
\end{equation}
where $a_{h}$ and $\varepsilon$ are measured parameters.

We add a systematic uncertainty of $0.1$ mag in quadrature to the fitted distance modulus to account for variations arising from the choice of isochrone library \citep{drlicawagner2015}.
The heliocentric distance is calculated from the distance modulus.
The uncertainty on the heliocentric distance includes the added systematic uncertainty in the distance modulus.
The absolute $V$-band magnitude, $M_V$, is calculated following the procedure described in \citet{martin2008}.
The stellar mass $M_*$ is estimated assuming a mass-to-light ratio of $M_*/L_V = 2$.

As one of its outputs, \texttt{ugali} assigns membership probabilities to every star in the field by comparing the spatial positions and photometric properties of each star to a model comprised of the satellite system and a spatially uniform population of contaminants (i.e., foreground stars and mis-classified background galaxies).
The number of detected stars, $N_*$, is defined as the sum of the membership probabilities.

Figure~\ref{fig:gr_6panel} shows a set of six diagnostic plots characterizing the objects in the Virgo~III field.
The top-left panel shows a smoothed isochrone-selected stellar density map with the overdensity in the center corresponding to Virgo~III.
The top-middle panel shows the spatial distribution of isochrone-selected stars (gray) with colored points indicating stars with ugali membership probability $>10\%$.
The top-right panel shows a $g$ vs.\ $g-r$ color--magnitude diagram (CMD) of all stars within a 2\,arcmin radius of the fitted centroid coordinates, and an isochrone with the best-fit parameters derived from the \texttt{ugali} MCMC fit.

The magnitude-dependent width of the isochrone filter is determined by two components.
The first component is a fit to the median uncertainty in $g-r$ color ($\sigma_{g-r}$) of stars within the stellar locus ($-0.5 < g-r < 1.1$) as a function of $g$-band magnitude using an exponential model of the form $\sigma_{g-r} = A + B\cdot10^{0.4(g -C)}$.
The second component accounts for the object-to-object scatter in $\sigma_{g-r}$ at fixed magnitude, which we quantify as the $84\mathrm{th} - 50\mathrm{th}$ percentile width of the $\sigma_{g-r}$ distribution in bins of $g$-band magnitude, and fit with an exponential model of the same form.
The total filter width is then calculated by adding these two components in quadrature.

The bottom-left panel presents an isochrone-selected smoothed galaxy density map, which serves as a diagnostic for identifying issues with star--galaxy classification and other imaging artifacts.
The bottom-middle panel shows the spatial distribution of point sources that pass and fail the isochrone selection filter.
The bottom-right panel shows a Hess diagram created from the isochrone-filtered stars selected within $4$\,arcmin of the centroid of Virgo~III compared to contaminants from a background annulus from $20$\,arcmin to $20.4$\,arcmin.

Similar plots using the $g-i$ and $r-i$ colors are shown in Appendix~\ref{app:isochrones} and are used to select the final high-purity member sample discussed in Section~\ref{sec:final_members}.
\vspace{10mm}
\subsection{Comparison with HSC results }
\label{sec:comparison}

\begin{figure*}[t!]
    \centering
    \includegraphics[width=0.95\textwidth]{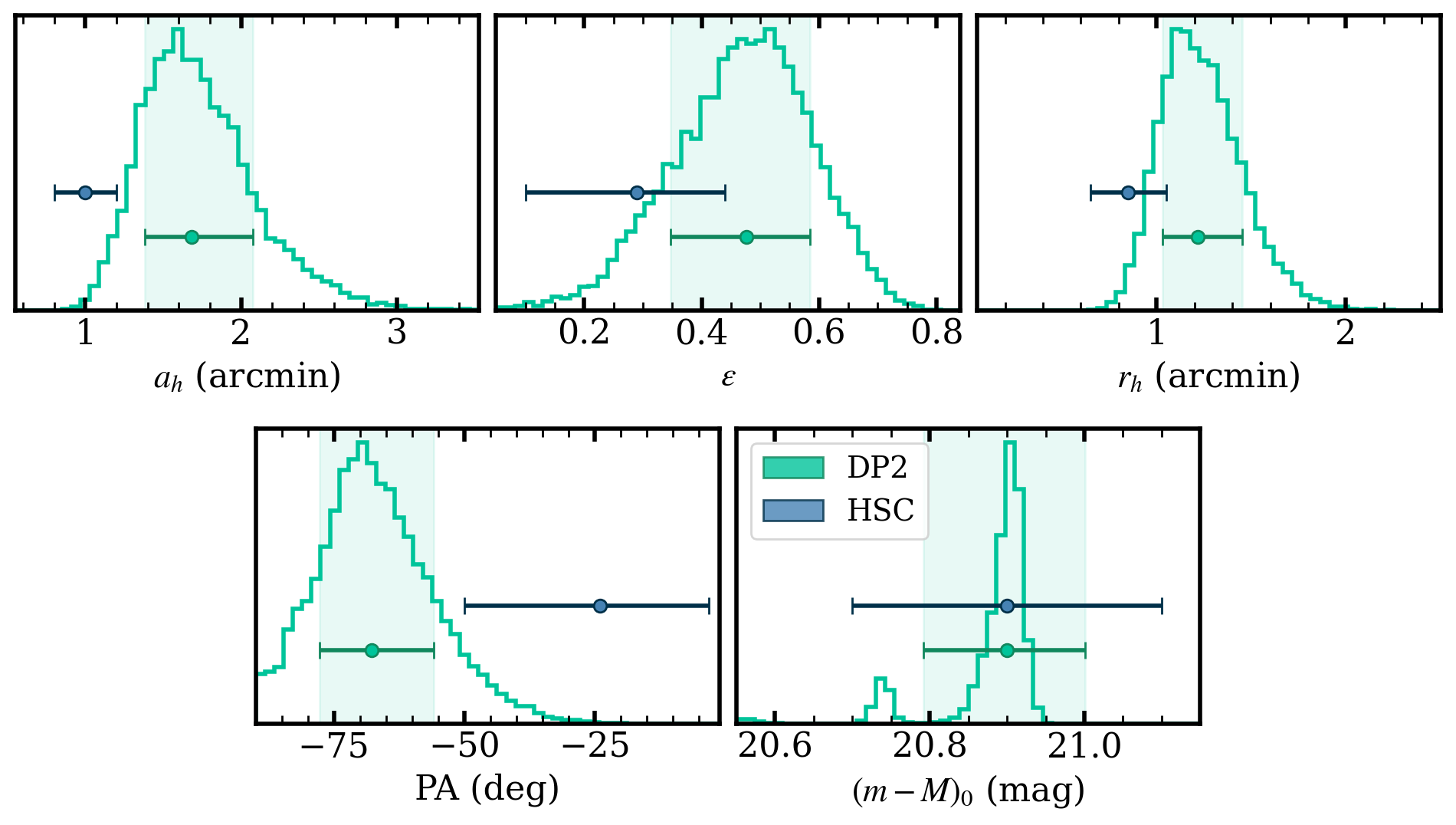}
    \caption{
    Comparison between our best-fit morphological parameters obtained from Rubin DP2 using \texttt{ugali} and the HSC measurements from \citet{homma2024}.
    The \texttt{ugali} marginalized posteriors are shown in green, along with a green point with error bars indicating the 50th percentile value and 1$\sigma$ uncertainties.
    The dark blue points show the best-fit HSC values with the reported uncertainties.
    Our fitted morphological parameters are broadly consistent with the HSC measurements.}
    
    \label{fig:comparison}
\end{figure*}
In this section, we compare our best-fit parameters of Virgo~III with those measured by \citet{homma2024}. 
Figure~\ref{fig:comparison} shows the marginalized posteriors of the angular semi-major axis $a_h$, ellipticity $\varepsilon$, azimuthally averaged angular half-light radius $r_h$, position angle, and distance modulus $(m-M)_0$ estimated with \texttt{ugali}. 
The green points with error bars and shaded bands denote the 50th, 16th, and 84th percentile values respectively, and the blue points denote the measured values and uncertainties from \citet{homma2024}.

Our DP2 measurements are mostly consistent with the previous HSC measurements to within their uncertainties, though we obtain a slightly different morphology, which we attribute to the greater depth of the DP2 data. 
The HSC measurements use stars with $i \lesssim 24.5$\,mag, while our sample extends to $g, r \sim 26$\,mag, with $\sim62.5\%$ of our high-purity member sample  (see next section) being fainter than $i = 24.5$\,mag. The \texttt{ugali} estimate of $N_{\rm obs} = \satnstar$ member stars is more than four times larger than the $N_* = 25^{+5}_{-4}$ reported by \citet{homma2024}, reflecting the greater depth of DP2.

We measure a distance modulus of $(m-M)_0 = \satmod$, where the uncertainty includes a systematic contribution of $\SI{0.1}{\mag}$ added in quadrature, corresponding to a heliocentric distance of $D_\odot = \satheliodist$ (also including the $\SI{0.1}{\mag}$ systematic), which agrees with $D_\odot = 151^{+15}_{-13}\,\si{\kpc}$ from \citet{homma2024}.
As noted by \citet{homma2024}, the distance estimate is strongly driven by the prominent horizontal branch (HB) associated with Virgo~III. We find that fitting the distance independently of the HB yields a distance modulus that is slightly closer ($(m-M)_0 \sim 20.65$). We note that the HSC discovery paper used four PARSEC isochrones with fixed ages and metallicities, whereas we use a finer isochrone grid with a broader range of ages and metallicities. 
Our measured absolute $V$-band magnitude of $M_V = \satmv$ is also consistent with the HSC measurement of $M_V = -2.69^{+0.45}_{-0.56}$. 

We measure a slightly larger ellipticity of $\varepsilon = \satell$, compared to $\varepsilon = 0.29^{+0.15}_{-0.19}$ from \citet{homma2024}, though these are consistent within $1\sigma$.
On the other hand, we find that our measured angular semi-major axis length and position angle do not agree with previous measurements. Our semi-major axis length $a_h = \satah$ is significantly larger than the previously measured value of $a_h = 1.0^{+0.2}_{-0.2}\,\si{\arcmin}$ and our position angle is $\rm{PA} = \satpa$ as compared to $\rm{PA} = -24^{+21}_{-26}\,\si{\deg}$.
Our morphological fit suggests that Virgo~III is more elongated and elliptical than the discovery results. 
We attribute this difference to the presence of an overdensity of isochrone-selected stars south of the centroid of Virgo~III which can be seen in the smoothed stellar density maps (top-left panel) of Figures~\ref{fig:gr_6panel}, \ref{fig:gi_6panel} and~\ref{fig:ri_6panel} around ($\Delta$R.A., $\Delta$Decl.) $\sim(0, -0.06)$.
This extended feature is not seen in the original discovery results, as the bulk of the isochrone-selected stars in this region are faint ($\gtrsim 25\,\si{\mag}$) and would have been below the magnitude limit of \citet{homma2024}.

\begin{figure*}[t!]
    \centering
    \includegraphics[width=0.95\textwidth]{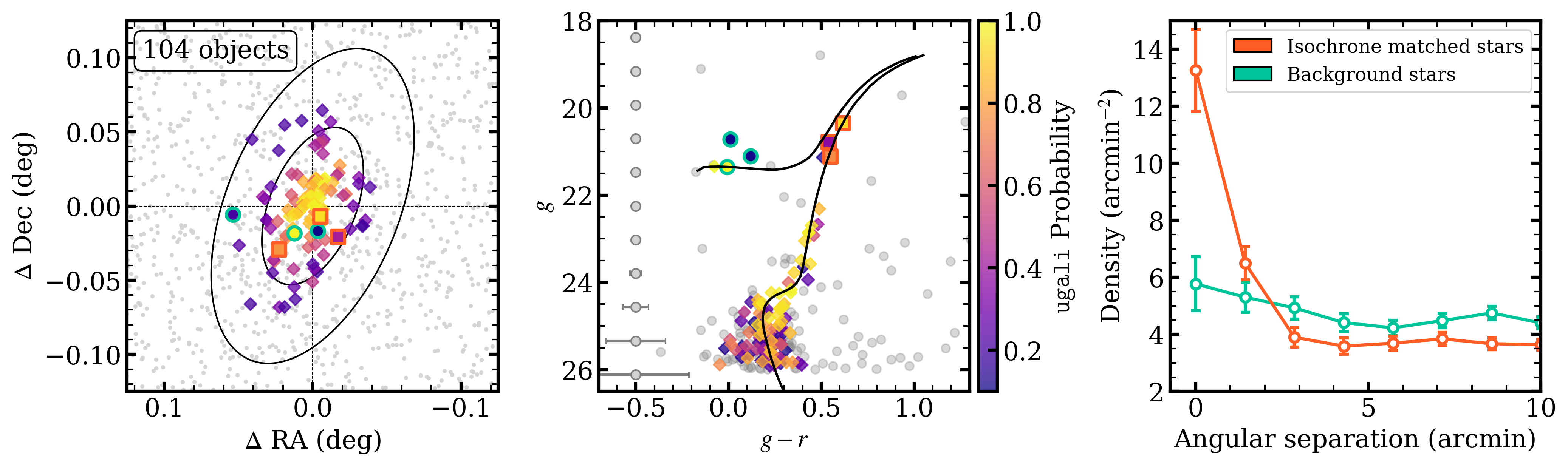}
    \caption{The high-purity Virgo~III member sample defined as stars selected with $\texttt{refExtendedness} <0.5$, \texttt{ugali} membership probability $> 10\%$, colors consistent with $g-r$, $r-i$, \& $g-i$ isochrones, and $|\mu_{\alpha*}| < 3\,\SI{}{\mas\per\year}$ \& $|\mu_\delta| < 3\,\SI{}{\mas\per\year}$ (if available).
    The left panel shows the spatial distribution of these member stars.
    Member stars are shown as diamonds colored by their \texttt{ugali} membership probability.
    The three known RR Lyrae are shown as circles with cyan outlines.
    RGB stars with \textit{Gaia} DR3 matches are shown as squares with orange outlines.
    Background stars are shown in gray.
    The middle panel shows the color--magnitude diagram (CMD), and the right panel shows the radial density profile of isochrone-selected (orange) and background (cyan) stars measured in circular annuli centered on Virgo~III. 
    The error bars represent the Poisson uncertainty in each annulus.
    }
    \label{fig:gr_final}
\end{figure*}

Finally, for the physical size we measure an azimuthally-averaged angular half-light radius of $r_h = \satrh$ as compared to the previously measured value of $r_h = 0.84^{+0.20}_{-0.19}\,\si{\arcmin}$ and an azimuthally-averaged physical half-light radius of $r_{1/2} = \satphysrh$ as compared to $r_{1/2} = 37^{+10}_{-9}\,\si{\pc}$. 
Note that \citet{homma2024} only provide measurements for the half-light radius along the semi-major axis; we calculate the azimuthally averaged values using Equation~\eqref{eq:rh} where we propagate provided uncertainties in ellipticity and heliocentric distance.
The uncertainties of both of the azimuthally averaged half-light radius measurements (angular and physical) are discrepant at the $\sim2\sigma$ level.
We attribute this discrepancy to the extended feature seen in the deeper DP2 data, leading to a larger size measurement of Virgo~III.
To verify this, we refit the morphological parameters of Virgo~III with a magnitude limit of $i < 24.5$ and observe that the tension with the HSC measurements is resolved.
Overall, we measure Virgo~III to be a larger, more elliptical system at a similar distance and luminosity as compared to previous measurements.

\subsection{Final Membership Selection }
\label{sec:final_members}
To assemble a high-purity sample of candidate member stars, we perform our analysis in several different color--magnitude combinations.  Using an isochrone matched filter in the $g-r$ color selects the largest number of stars (189) within the filter, while using the $g-i$ color selects the fewest (120).
We conservatively narrow the member sample by only retaining stars with \texttt{ugali} membership probability $> 10\%$ and colors that are consistent with all three isochrone selections.
We also remove stars with \textit{Gaia} Data Release 3 \citep[DR3;][]{prusti2016, vallenari2023} proper motions $|\mu_{\alpha*}| > 3\,\SI{}{\mas\per\year}$ or $|\mu_\delta| > 3\,\SI{}{\mas\per\year}$, since we expect the proper motions of member stars to be clustered around $(\mu_{\alpha*}, \mu_\delta) \sim (0, 0)$\,\SI{}{\mas\per\year} at such a large heliocentric distance.
Here, $\mu_{\alpha*} \equiv \mu_\alpha \cos \delta$ and $\mu_\delta$ are the {\it Gaia} proper motion components in R.A.\ and Decl., respectively.

Figure~\ref{fig:gr_final} shows the spatial distribution (left panel) and CMD (middle panel) of the final high-purity sample of 104 candidate member stars.
Stars with available proper motions in \textit{Gaia} DR3 (matched with $\leq 1$\,arcsec separation) are shown as squares with orange outlines along the RGB.
Known RR Lyrae \citep{ngeow2024} are shown as circles with cyan outlines.
The right panel also shows the radial density profile of isochrone-selected stars (orange) measured in circular annuli centered on Virgo~III.
Background stars that are not selected by the isochrone matched filter are shown in cyan.
We note that we do not account for any magnitude spread in the HB, leading to a low membership probability being assigned to the two RR Lyrae that are slightly brighter than the HB model.

\section{Properties of Member Stars }
\label{sec:mem_properties}
\subsection{Gaia Data}
We associate stars in our high-purity Virgo~III member sample (Section~\ref{sec:morph}) with measurements from \textit{Gaia} DR3 \citep[][]{prusti2016, vallenari2023}.
We query all stars within 10\,arcmin of the centroid of Virgo~III that satisfy the following criteria from \citet{pace2022}.
A full five-parameter astrometric solution must have been fit to the source (\texttt{astrometric\_params\_solved} $> 3$).
Sources whose astrometric residuals are significantly larger than expected from the measurement uncertainties are removed via a cut on \texttt{astrometric\_excess\_noise\_sig} $< 2$.
The renormalized unit weight error \citep[\texttt{ruwe};][]{lindegren2021} is required to be $\texttt{ruwe} < 1.3$, excluding sources for which the single-star astrometric model is a poor fit, often indicative of binarity or blending.
Blended or extended sources are further removed by requiring \texttt{ipd\_frac\_multi\_peak} $< 2$.
A minimum of 10 distinct visibility periods (\texttt{visibility\_periods\_used} $> 10$) is required to ensure reliable proper motion measurements.
Finally, sources with a parallax $\varpi$ inconsistent with zero at the $3\sigma$ level are excluded via the condition $\varpi - 3\sigma_\varpi < 0$, removing foreground star contamination while retaining distant sources whose true parallax is effectively zero. 
We then perform a 1\,arcsec spatial cross match to associated objects between {\it Gaia} DR3 and Rubin DP2.

\subsection{Proper Motions and Orbital Constraints}
\label{sec:prop_motions}
\begin{figure}[t!]
    \centering
    \includegraphics[width=0.95\columnwidth]{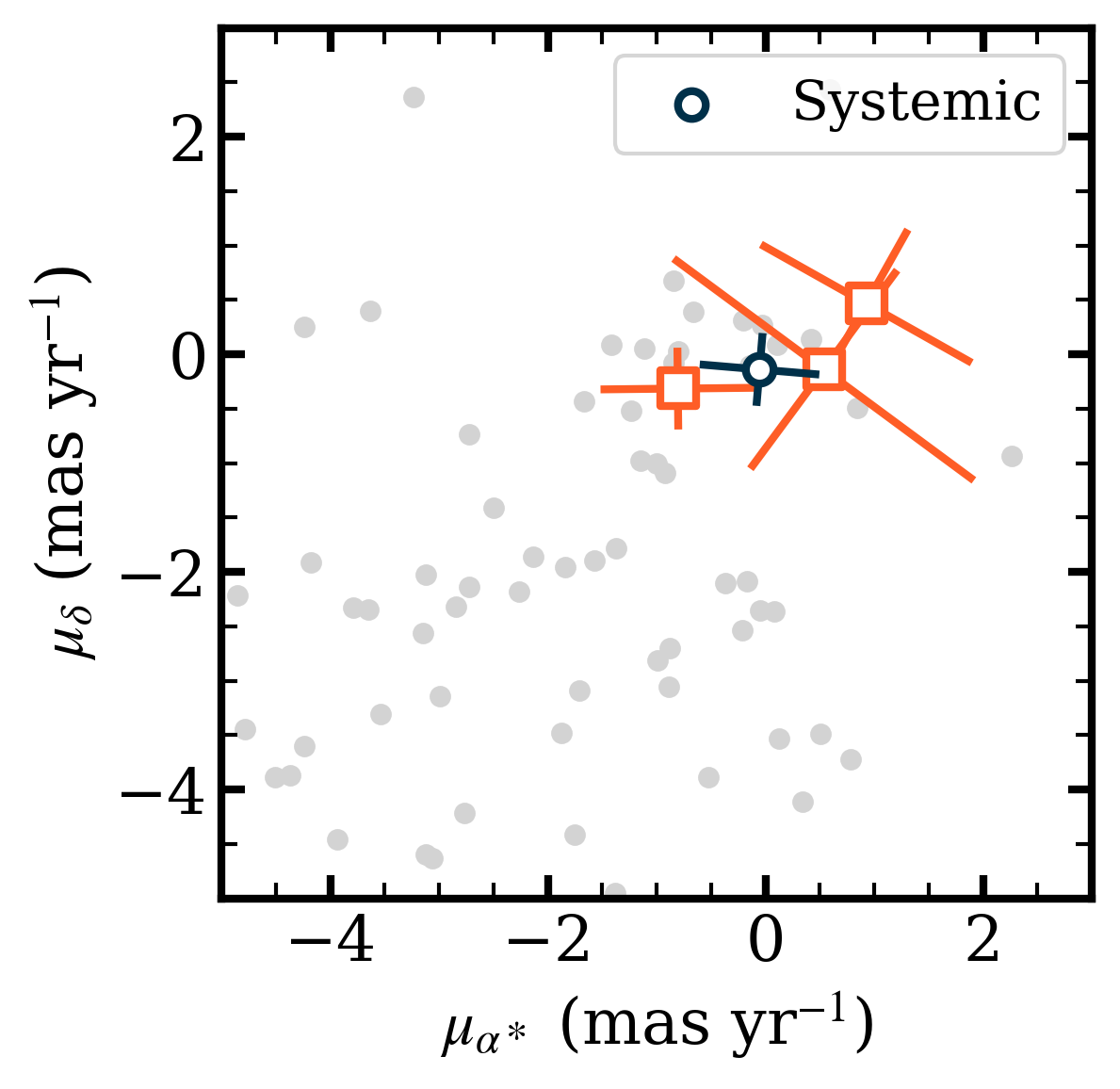}
    \caption{
    Proper motions of stars within a 10\,arcmin radius of the centroid of Virgo~III. 
    Background stars are plotted as gray circles, while stars belonging to the Virgo~III high-purity member sample are plotted as orange squares.
    The \texttt{ugali} membership probabilites from left to right are $P_{\texttt{ugali}} = 0.93, 0.36$ and $0.75$ (cf.\ Table~\ref{tab:pm}).
    The proper motion uncertainties include the covariance between the two coordinate components.
    The dark blue circle shows the systemic proper motion of the system.
    The systemic proper motion is close to $(0, 0)$\,\SI{}{\mas\per\year} as is expected for a distant Milky Way halo system. 
    }
    \label{fig:pm}
\end{figure}

Figure~\ref{fig:pm} shows the proper motions of the stars within a 10\,arcmin radius of Virgo~III.
We find proper motions from $Gaia$ DR3 for three RGB stars from the final high-purity member sample.
The proper motions of these member stars are clustered around $(\mu_{\alpha*}, \mu_\delta) \sim (0, 0)$\,\SI{}{\mas\per\year} as is expected for a distant gravitationally bound Milky Way halo system.
We calculate a systemic proper motion by using \texttt{emcee} to sample the posterior distributions of $\mu_{\alpha*}$ and $\mu_\delta$ accounting for the covariance between the two following \citet{pace2019}. 

\begin{figure*}[t!]
    \centering
    \includegraphics[width=0.95\textwidth]{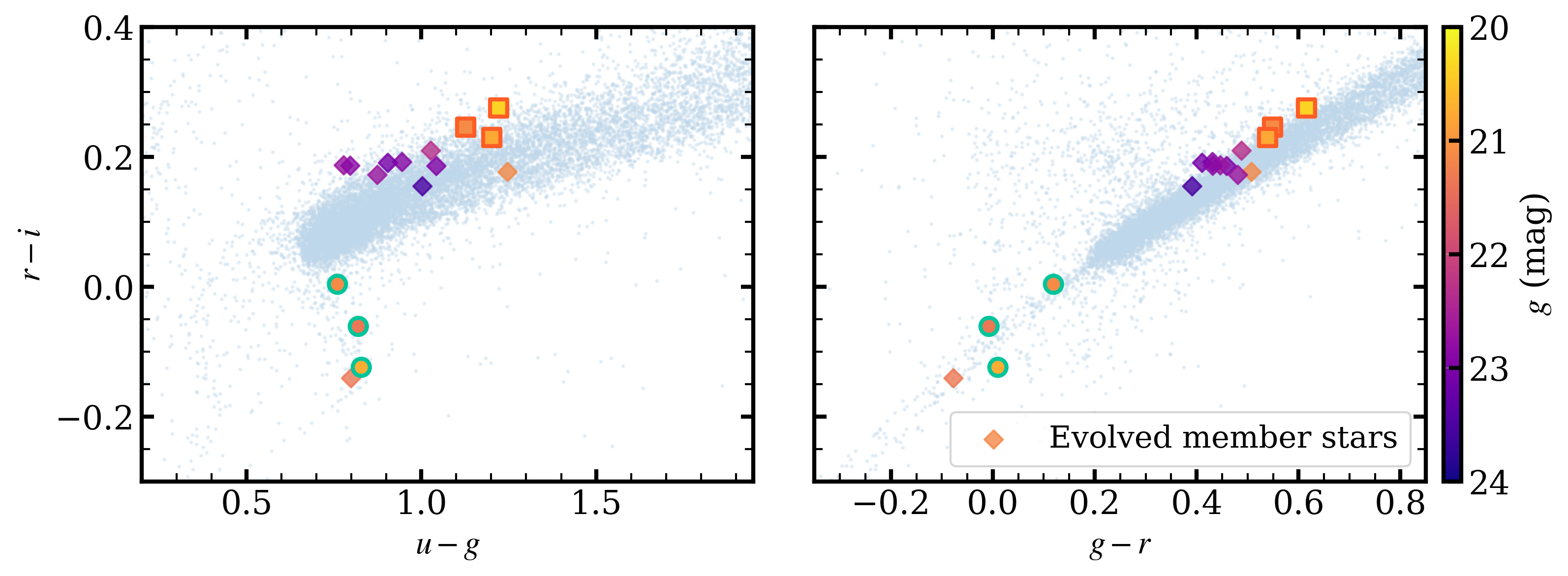}
    \caption{
    Distributions of the evolved, post-main sequence member stars of Virgo~III in color--color space.
    Virgo~III member stars are plotted as diamonds colored by their $g$-band magnitudes.
     Known RR Lyrae are plotted as circles with cyan outlines.
     RGB stars with matches to \textit{Gaia} DR3 are shown as squares with orange outlines.
    In both diagrams, the light blue stellar loci are composed of stars with $g$-band $\rm SNR > 100$.
    The left panel shows an $r-i$ vs.\ $u-g$ color--color diagram. 
    The HB stars lie below $r-i < 0.1$, forming a hook-like spur at the tip of the main stellar locus.
    The right panel shows an $r-i$ vs.\ $g-r$ color--color diagram. 
    }
    \label{fig:color-color}
\end{figure*}

\begin{deluxetable}{ccc}[h!]
\tablecaption{Rubin \texttt{Object} catalog ID, $Gaia$ source ID, and \texttt{ugali} membership probabilities for high-probability member stars with available proper motions.
\label{tab:pm}}
\tablehead{
    \colhead{\texttt{Object} ID} &
    \colhead{$Gaia$ Source ID} &
    \colhead{$P_\texttt{ugali}$}
}
\startdata
789844623173377982 & 3707878256281422976 & 0.93 \\
789844623173376690 & 3707877431647316480 & 0.75 \\
789844623173377220 & 3707878114546739840 & 0.36 \\
\enddata
\end{deluxetable}

The $Gaia$ source IDs and \texttt{ugali} membership probabilities of the three stars are shown in Table~\ref{tab:pm}.
The systemic proper motion is found to be $\mu_{\alpha*}^{\rm sys} = -0.06 \pm 0.51\,\SI{}{\mas\per\year}$ and $\mu_{\delta}^{\rm sys} = -0.14 \pm 0.30\,\SI{}{\mas\per\year}$.
These uncertainties of $\sigma_{\mu_{\alpha*}} = \SI{0.51}{\mas\per\year}$ 
and $\sigma_{\mu_\delta} = \SI{0.30}{\mas\per\year}$ correspond to tangential velocity uncertainties of $\sim$\SI{365}{\kilo\m\per\s} and $\sim$\SI{215}{\kilo\m\per\s}, respectively, at a heliocentric distance of $D_\odot \sim \satheliodist$, 
and thus are comparable, if not greater, than the escape velocity at that distance \citep[\SI{\sim 250}{\kilo\m\per\s};][]{bovy2015}.
As a result, any attempt to infer the range of line-of-sight velocities that would lead to bound orbits is susceptible to a large fraction of orbits being classified as unbound due to the uncertainties in the systemic proper motion alone, preventing us from drawing any meaningful conclusions about the orbit of Virgo~III.

\subsection{Photometric Metallicity }
\label{sec:colorcolor}
In this section, we explore the distribution of high-purity candidate member stars in various color--color spaces.
Figure~\ref{fig:color-color} shows the distributions of member stars in $u-g$ vs.\ $r-i$ (left panel) and $g-r$ vs.\ $r-i$ (right panel).
The member stars are colored by their $g$-band magnitudes, while the background stellar locus (within $\sim2$\,deg of Virgo~III) is plotted in light blue.
The HB stars lie on a tight, hook-like spur around $u-g\sim 0.75$ and $r-i < 0.1$.
This is due to the shifting of the Balmer break in hot stars between the $u$ and $g$ bands causing a plateau in $u-g$ color, while the $r-i$ color continues to become steadily bluer as the spectra of these stars are dominated by the Rayleigh-Jeans tail in the $r$ and $i$ bands.

\begin{figure}[h!]
    \centering
    \includegraphics[width=0.95\columnwidth]{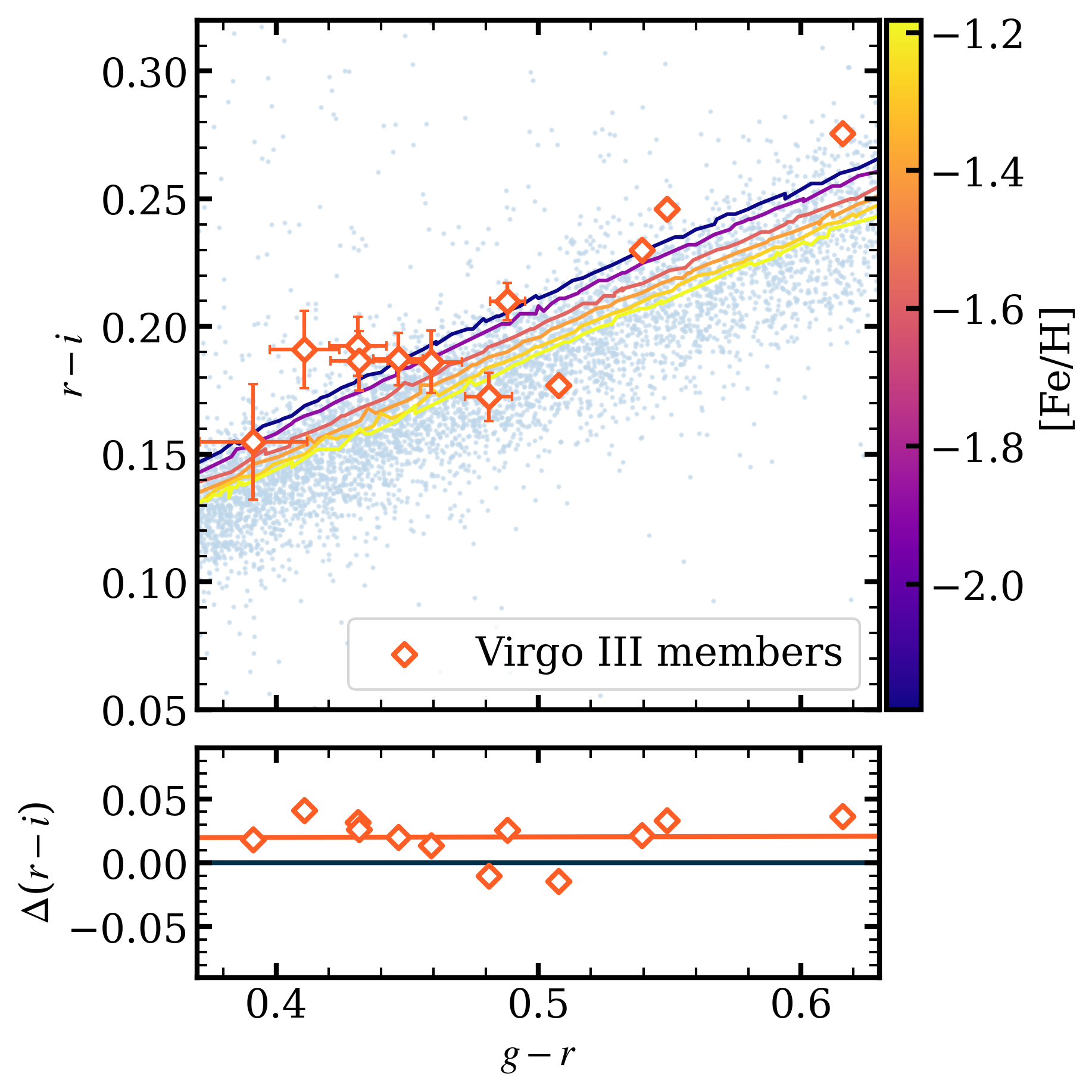}
    \caption{
    The top-panel shows a $g-r$ vs.\ $r-i$ color--color diagram for stars in the Virgo~III field. 
    The background stellar locus is shown as light blue points. 
    Virgo~III member stars that are brighter than $g < \SI{23.75}{\mag}$ and $r < \SI{23.75}{\mag}$ are shown as unfilled orange diamonds.
    The colored lines show isochrones with a fixed age of \SI{13}{Gyr} and varying metallicity.
    As the metallicity decreases, the $r-i$ color of the isochrones shifts redward for a given $g-r$ color.
    The bottom panel shows a median-subtracted residual plot.
    The dark blue line is the median of the background stellar locus and the orange line is the residual of the line of best fit to Virgo~III member stars that lie between $0.35 < g-r < 0.7$.
    The high-probability member stars are systematically offset above the stellar locus by $\sim$\SI{20}{mmag}, indicating a metal poor population.}
    \label{fig:metal_poor}
\end{figure}

\citet{li2019} used spectroscopy from $S^5$ to show that DES photometry was sufficient to separate metal-rich and metal-poor evolved, post-MSTO stars in $g-r$ vs.\ $r-i$ and $g-r$ vs.\ $r-z$ color--color spaces.
The metal-poor stars were observed to lie towards the top of the stellar locus between $0.4 \leq g-r \leq 0.8$.

We find that the DP2 photometry is also of sufficient quality to separate metal-poor and metal-rich stars in color--color space, consistent with the trend observed above.
Figure~\ref{fig:metal_poor} shows a $g-r$ vs.\ $r-i$ color--color diagram overplotted with colors of stars from isochrones of different metallicities at the same age (\SI{13}{Gyr}).
The light blue points show the background stellar locus.
The unfilled orange diamonds show stars that are a part of the final member sample of Virgo~III.
The bottom panel shows the residual between a line fitted to Virgo~III member stars in orange and the median of the background stellar locus, showing a systematic offset of approximately $+$\SI{20}{mmag} for stars as faint as \SI{23.75}{\mag}.
We expect this separation to improve in subsequent LSST data releases as they are expected to further reduce the scatter in the foreground population, thus separating the metal poor member population to a higher degree and enabling photometric metallicity estimates for large samples of faint satellite systems in the future.

\begin{figure*}[t!]
    \centering
    \gridline{
        \includegraphics[width=0.3\textwidth]{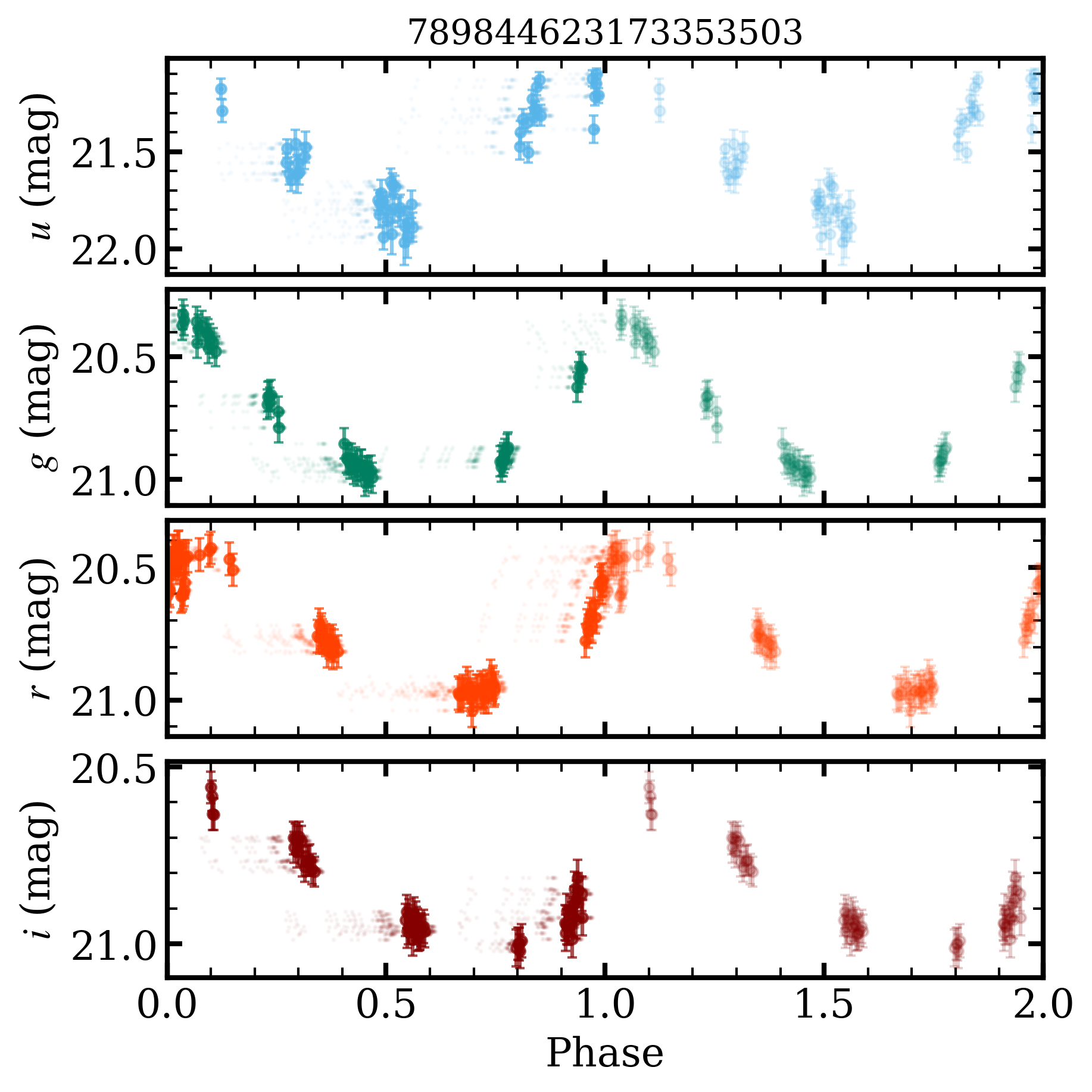}
        \includegraphics[width=0.3\textwidth]{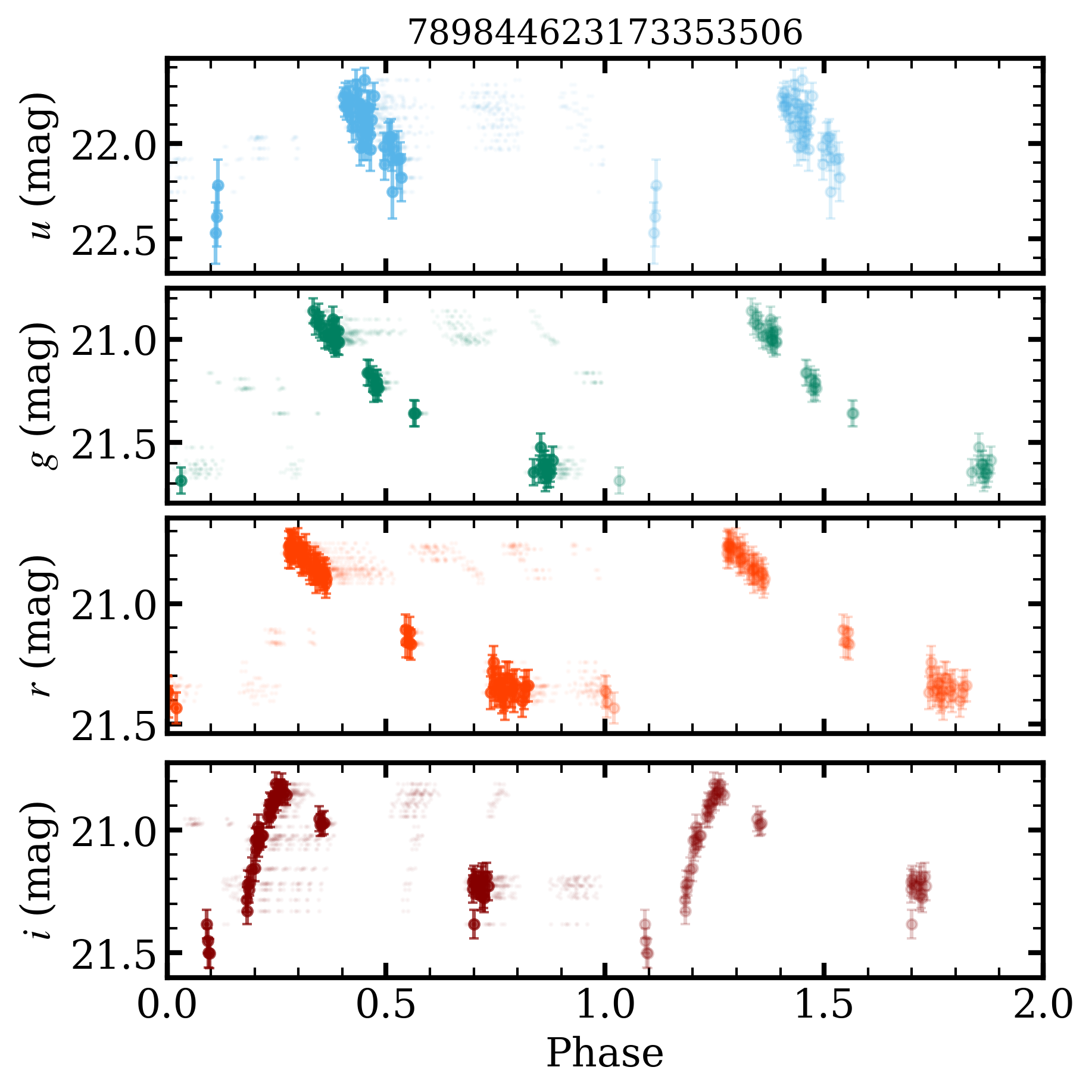}
        \includegraphics[width=0.3\textwidth]{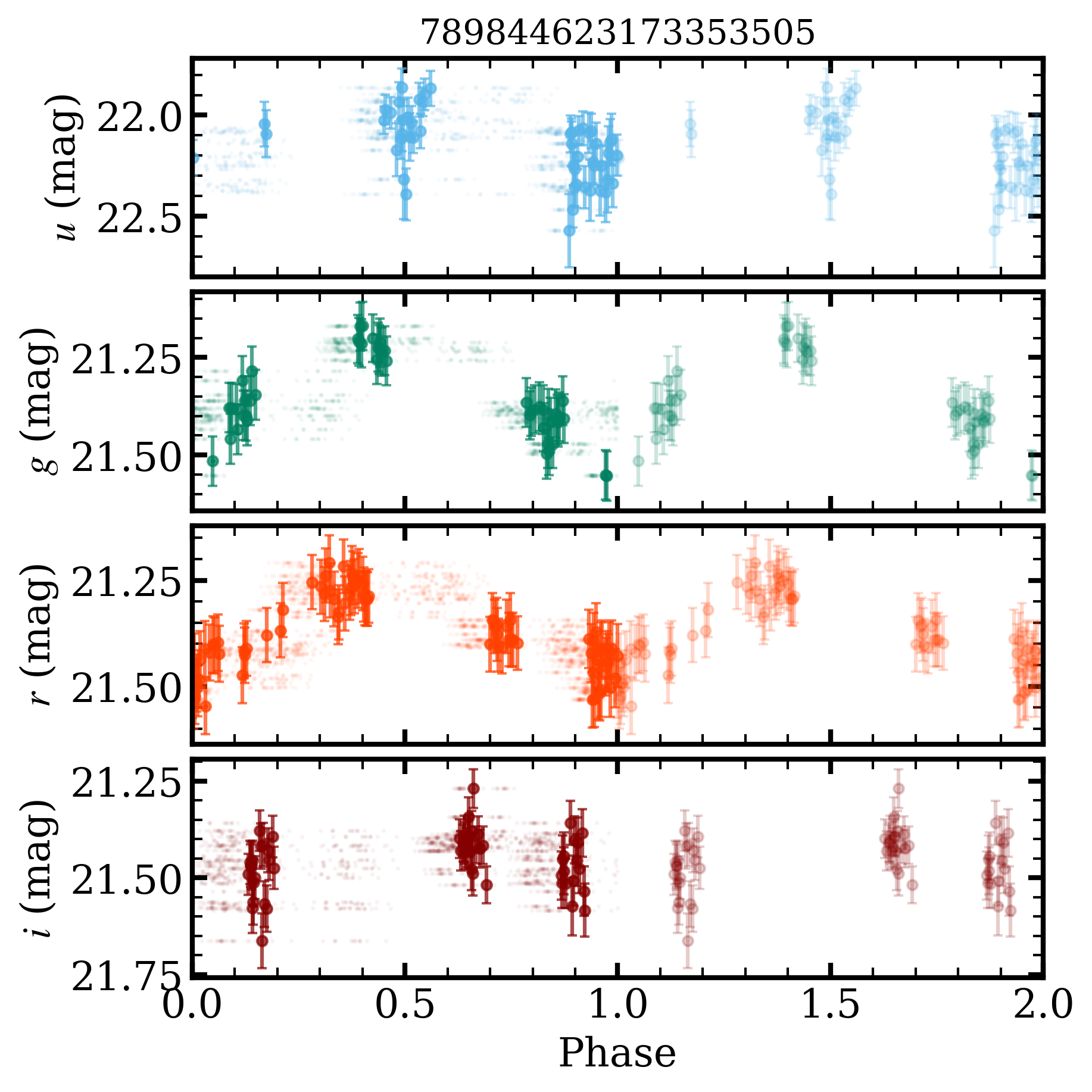}
    }
    \gridline{
        \includegraphics[width=0.3\textwidth]{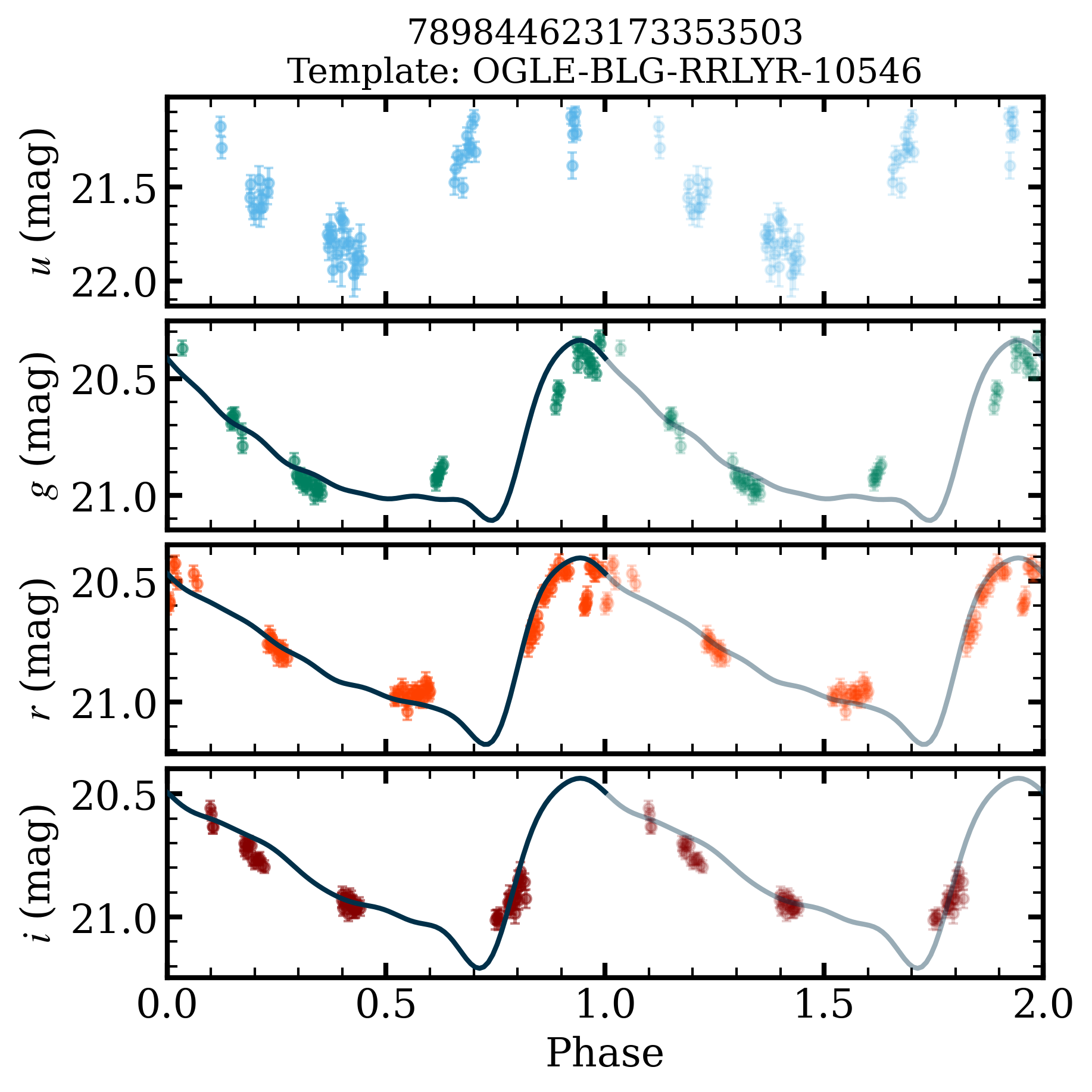}
        \includegraphics[width=0.3\textwidth]{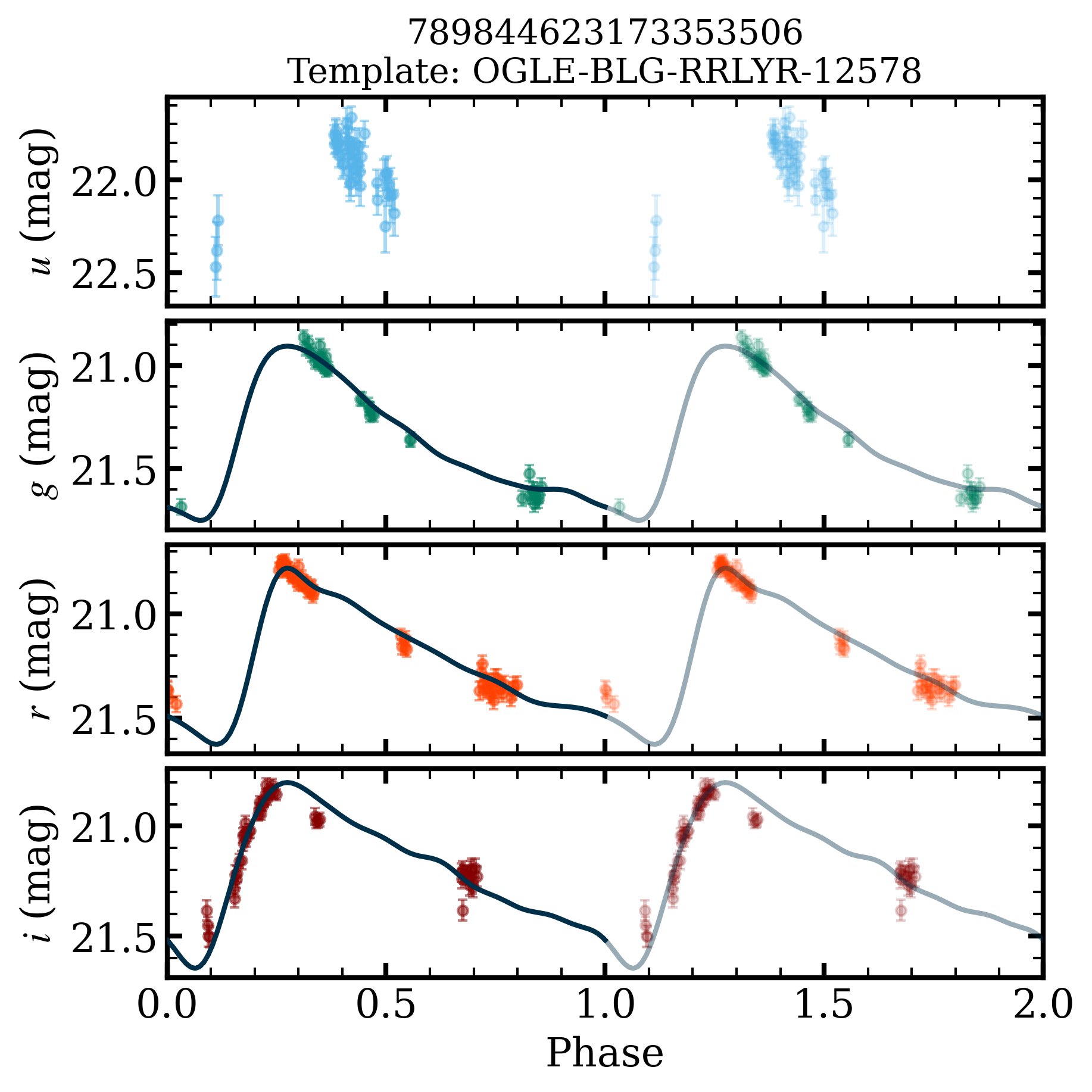}
        \includegraphics[width=0.3\textwidth]{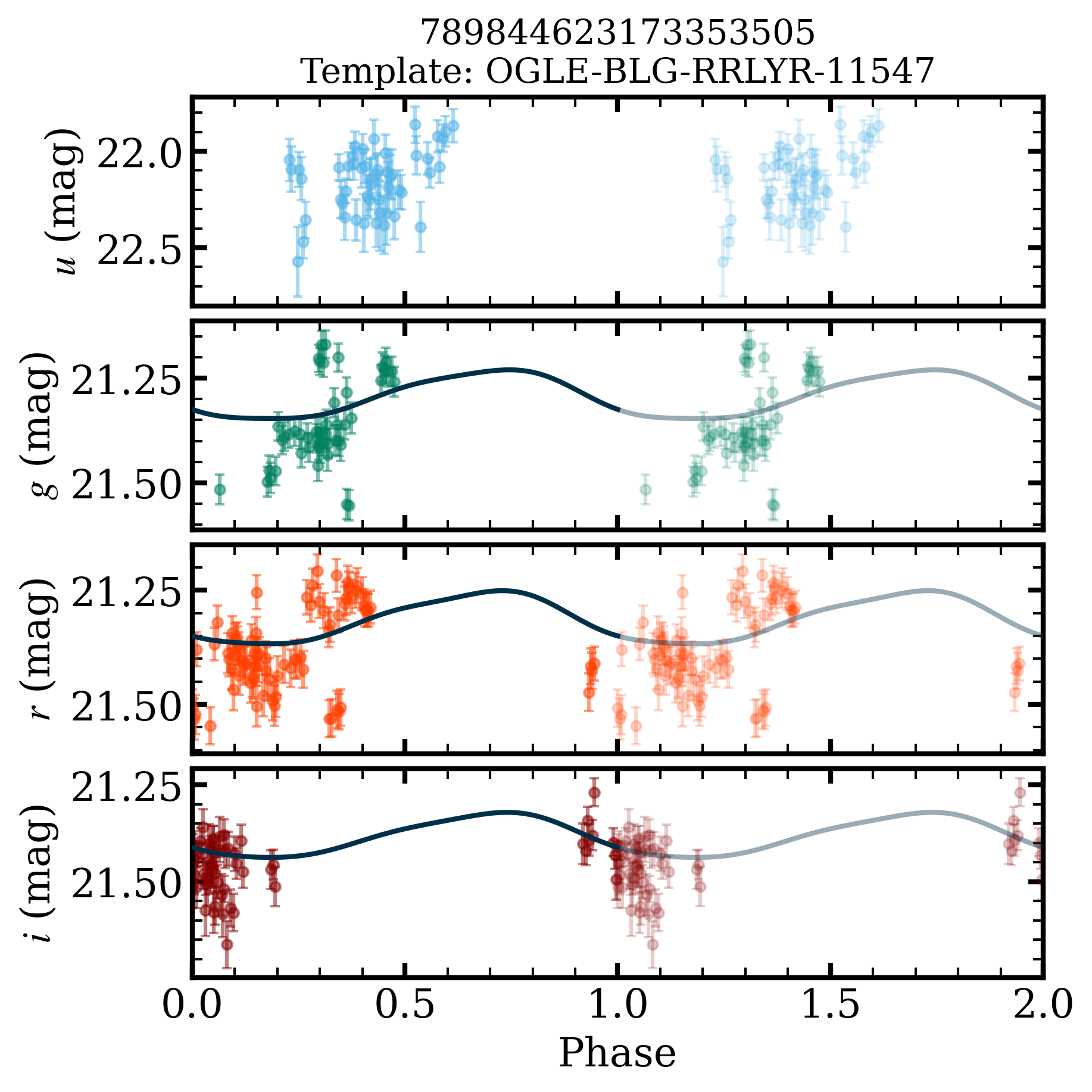}
    }
    \caption{
    Bootstrapped phase-folded light curves of the three known RR Lyrae in Virgo~III. 
    The top row shows phase-folded light curves from the period search algorithm.
    For each star, the $ugri$ light curves are shown from top to bottom. 
    Each panel is labelled by its \texttt{diaObjectId} value.
    The bottom row shows phase-folded light curves from the template fitting procedure.
    Each panel is labelled by its \texttt{diaObjectId} and the best-fit template name (from the template library of \citealt{baezavillagra2025}).
    The solid line in the $gri$ panels shows the best-fit RR Lyrae template.
    The left and center panels in each row show the two RRab stars, and the right panel shows the RRc star.
    \label{fig:rrl}}
\end{figure*}

\section{RR Lyrae }
\label{sec:rrl}
\citet{ngeow2024} discovered three RR Lyrae stars in Virgo~III, of which two were identified as Type ab (RRab) and one as Type c (RRc).
The two RRab stars were labeled `V1' and `V2', while the RRc star was labeled `V3'.
In this section, we refer to these stars as `RRL~1', `RRL~2', and `RRL~3', respectively. 
The DP2 \texttt{diaObjectId} values for each star are reported in Table~\ref{tab:rrl}.

Measurements of these objects are available in the DP2 \texttt{DiaSource} and \texttt{ForcedSourceOnDiaObject} tables.
The former contains measurements of variable object detections on single-visit difference images, while the latter contains forced-photometry measurements on the single-visit images.
We opt to use the \texttt{ForcedSourceOnDiaObject} tables, since they contain more measurements than the \texttt{DiaSource} tables.
We use the \texttt{pycycle}\footnote{\url{https://github.com/psferguson/pycycle}} Python package to estimate periods of known RR Lyrae using two methods --- template fitting and period searching.
We then use the fitted periods to measure distance moduli using the Period-Luminosity-Metallicity (PLZ) and Period-Wesenheit-Metallicity (PWZ) relations.
We also search for other variable stars within a 0.5\,deg radius of Virgo~III, but we find no objects that show significant periodicity.

\subsection{Light Curve Processing}
Before fitting the period of the RR Lyrae with either method, we perform the following processing steps on the data obtained from the \texttt{ForcedSourceOnDiaObject} catalogs.

We remove observations that are flagged with
\texttt{invalidPsfFlag},
\texttt{pixelFlags\_nodata}, \texttt{pixelFlags\_bad}, \texttt{pixelFlags\_cr}, \texttt{pixelFlags\_edge}, \texttt{pixelFlags\_saturated},
\texttt{pixelFlags\_suspect} or \texttt{psfFlux\_flag}.
We also find that the reported photometric uncertainties from the \texttt{ForcedSourceOnDiaObject} must be inflated to accurately represent the scatter in the flux of static sources across epochs.
For each RR Lyrae, we estimate a systematic calibration uncertainty in each band as the standard deviation of the distribution of zeropoints for images in which the star was observed.
These uncertainty estimates amount to \SI{0.03}{mag} in $g$ and $r$, and \SI{0.02}{mag} in $i$.
We add this systematic uncertainty in quadrature with the reported uncertainties.

\subsection{Period Search}
The \texttt{pycycle} package uses a hybrid Lomb-Scargle and Lafler-Kinman period-finding algorithm based on the framework of \citet{saha2017}.
The best-fit periods are shown in Table~\ref{tab:rrl}.
We estimate uncertainties on the period by bootstrapping the lightcurve with $N=200$ resamplings for each RR Lyrae star, and report the 16th and 84th percentiles of the bootstrapped distribution.
The best-fit periods obtained from the period search algorithms are consistent with previous measurements from \citet{ngeow2024}.

\subsection{Template Fitting }
\label{sec:rrl-fit}
We use multiband RR Lyrae templates made for the DES $griz$ bands from \citet{baezavillagra2025}.
This template library consists of 136 RRab templates and 144 RRc templates.
We use the \texttt{apply\_des\_to\_lsst\_correction} function provided in  \texttt{pycycle} to transform the templates from the DECam to the LSSTCam bandpasses using empirical corrections derived from 26 RRab stars in the M49 field observed in Rubin DP1.\footnote{\url{https://github.com/psferguson/pycycle/blob/main/notebooks/fink\_rrl\_fit.ipynb}}
For these fits, we use only the $gri$ bands, since $u$-band templates are unavailable.

Figure~\ref{fig:rrl} shows the phase-folded light curves of each star in the $g$, $r$, and $i$ bands from top to bottom, with the best-fit template shown as a solid dark blue line.
The left and center panels show the light curves of the RRab stars, and the right panel shows the light curve of the RRc star.
Table~\ref{tab:rrl} shows the best-fit template parameters for each star.
Uncertainties on the period, amplitude, and mean magnitudes are estimated by bootstrapping the light curve observations with $N = 200$ resamplings for each RR Lyrae, and are reported as the 16th and 84th percentiles of the resulting distributions.

Although the best-fit periods of the two RRab stars are consistent with the 
values measured by \citet{ngeow2024}, the best-fit period for the RRc star ($P = 0.491$\,days) differs from the previously reported value ($P = 0.434$\,days).
The median uncertainty (in each band) of the two RRab light curves is at the $\sim$\SI{0.05}{mag} level, while the median uncertainty of the RRc light curve is slightly larger at $\sim$\SI{0.06}{mag}.
The larger uncertainty, combined with the generally smaller light curve amplitude ($0.2$--$0.4$\,mags) of RRc stars relative to RRab stars ($0.5$--$1$\,mags), makes the template fits less sensitive to the true periodicity.

In the analysis presented in the subsequent sections, we choose to use the best-fit periods derived from the period search method, as these periods are independent of template-related systematics such as metallicity dependence and systematics that could arise due to the DES-to-LSST bandpass transformation.

\vspace{10mm}
\subsection{Period-Luminosity-Metallicity Relation}
\label{sec:rrl-plz}
We use the derived PLZ relations for LSST bandpasses from Table~4 of \citet{marconi2022} to estimate distance moduli of the three RR Lyrae.
Equation~\eqref{eq:plz} shows the functional form of the PLZ relations.
Here, $M$ is the band-specific absolute magnitude, $P$ is the fitted period (from the period search method), and $\rm [Fe/H]$ is the metallicity of the RR Lyrae. 
As we do not have spectroscopic metallicity measurements for these RR Lyrae, we need to assume some metallicity. 
We generally find that the photometric metallicity from isochrone fitting is slightly higher than spectroscopic metallicity. 
Based on the mass-metallicity relationship, we assume ${\rm [Fe/H]} = \SI{-2.5}{\dex}$ ($Z = 0.00005$) \citep{pace2025}.
We find that our results are fairly insensitive to this choice: adopting instead the median of the fitted systemic isochrone metallicity does not significantly alter our conclusions.
The $\alpha, \beta, \gamma$ parameters are derived coefficients from \citet{marconi2022},
\begin{equation}
\label{eq:plz}
    M = \alpha + \beta \log_{10}{P} + \gamma\,{\rm [Fe/H]}.
\end{equation}
We use the $r$ and $i$ band fundamental mode coefficients to fit RRL~1 \& RRL~2, and the first-overtone coefficients to fit RRL~3.
The resulting distance moduli are shown in Table~\ref{tab:rrl}.
We add the RMS residual in the fitted relation (column labelled `$\sigma$' in Table~4 of \citealt{marconi2022}) in quadrature with the uncertainty propagated from the fitted coefficients, period, and metallicity.
These distance moduli ($r$-band) correspond to heliocentric distances of
$D_\odot = 114 \pm 8$\,kpc, $138 \pm 10$\,kpc, and $188 \pm 10$\,kpc for RRL~1, RRL~2, and RRL~3, respectively.
The distance for RRL~2 is statistically consistent with the heliocentric distance from the best-fit isochrone ($D_\odot = \satheliodist$), while the distances for RRL~1 and RRL~3 differ by $> 3\sigma$ in either direction.

\subsection{Period-Wesenheit-Metallicity Relation}
\label{sec:rrl-pwz}
We also use the derived PWZ relations from Table~4 of \citet{marconi2022} to estimate distance moduli.
Equation~\eqref{eq:pwz_w} shows the functional form of the Wesenheit magnitudes, while Equation~\eqref{eq:pwz} shows the PWZ relation,
\begin{align}
    W_r &= r - 2.796\,(g - r), \label{eq:pwz_w} \\
    W_i &= i - 1.287\,(g - i), \nonumber  \\[10pt]
    M_W &= \alpha_W + \beta_W \log_{10}{P} + \gamma_W\,{\rm [Fe/H]}. 
    \label{eq:pwz}
\end{align}
The distance modulus is calculated as the difference $\mu = W - M_W$ in the $r$ and $i$ bands.
The distance moduli estimated from this method are also shown in Table~\ref{tab:rrl}.
These distance moduli ($r$-band) correspond to
heliocentric distances of $D_\odot = 134 \pm 6$\,kpc, $131 \pm 7$\,kpc, and $200 \pm 8$\,kpc for RRL~1, RRL~2, and RRL~3, respectively.
Repeating the procedure from Section~\ref{sec:rrl-plz}, we have added the RMS residual in the fitted relation in quadrature with the uncertainty propagated from the fitted coefficients, period, and metallicity.
The distance of RRL~1 differs from the heliocentric distance from the best-fit isochrone ($D_\odot = \satheliodist$) at the $\sim 3\sigma$ level, while the distance estimated for RRL~3 is $\sim 6\sigma$ discrepant in the opposite direction.

\begin{deluxetable*}{lccc}[t!]
\tablecaption{Properties of the three RR Lyrae in Virgo~III, containing parameters derived from template fitting and period searching.
The $r$ and $i$ band distance moduli are discussed in Section~\ref{sec:rrl-plz} and Section~\ref{sec:rrl-pwz}.
\label{tab:rrl}}
\tablehead{
    \colhead{Property} & 
    \colhead{RRL 1} & 
    \colhead{RRL 2} & 
    \colhead{RRL 3}
}
\startdata
\texttt{diaObjectId} & $789844623173353503$ & $789844623173353506$ & $789844623173353505$ \\
Type & AB & AB & C \\
\noalign{\vskip 0.6ex}\hline\noalign{\vskip 0.6ex}
\multicolumn{2}{l}{\textbf{Template Fit Parameters}} \\
Template ID & 10546 & 12578 & 11547 \\
Period (days) & $0.6174^{+0.0012}_{-0.0029}$ & $0.6762^{+0.0018}_{-0.0002}$ & $0.4907^{+0.0014}_{-0.0014}$ \\
Amplitude (mag) & $0.77^{+0.02}_{-0.02}$ & $0.85^{+0.02}_{-0.01}$ & $-0.27^{+0.04}_{-0.06}$ \\
$\langle g \rangle$ (mag) & $20.34^{+0.02}_{-0.02}$ & $20.90^{+0.01}_{-0.02}$ & $21.38^{+0.02}_{-0.01}$ \\
$\langle r \rangle$ (mag) & $20.41^{+0.01}_{-0.02}$ & $20.78^{+0.01}_{-0.01}$ & $21.39^{+0.01}_{-0.01}$ \\
$\langle i \rangle$ (mag) & $20.44^{+0.02}_{-0.01}$ & $20.81^{+0.02}_{-0.02}$ & $21.45^{+0.01}_{-0.01}$ \\
\noalign{\vskip 0.6ex}\hline\noalign{\vskip 0.6ex}
\multicolumn{2}{l}{\textbf{Period Search Parameters}} \\
Period (days) & $0.6114^{+0.0027}_{-0.0009}$ & $0.6748^{+0.0484}_{-0.0024}$ & $0.4417^{+0.0021}_{-0.0050}$ \\
\noalign{\vskip 0.6ex}\hline\noalign{\vskip 0.6ex}
\multicolumn{2}{l}{\textbf{Distance Modulus from PLZ Relations}} \\
$\mu_r$ (mag) & $20.28 \pm 0.15$ & $20.70 \pm 0.16$ & $21.37 \pm 0.11$ \\
$\mu_i$ (mag) & $20.29 \pm 0.13$ & $20.73 \pm 0.14$ & $21.38 \pm 0.10$ \\
\noalign{\vskip 0.6ex}\hline\noalign{\vskip 0.6ex}
\multicolumn{2}{l}{\textbf{Distance Modulus from PWZ Relations}} \\
$\mu_r$ (mag) & $20.64 \pm 0.10$ & $20.58 \pm 0.12$ & $21.50 \pm 0.08$ \\
$\mu_i$ (mag) & $20.48 \pm 0.08$ & $20.71 \pm 0.11$ & $21.45 \pm 0.07$ \\
\hline
\enddata
\end{deluxetable*}
\begin{figure}[h!]
    \centering
    \includegraphics[width=0.95\columnwidth]{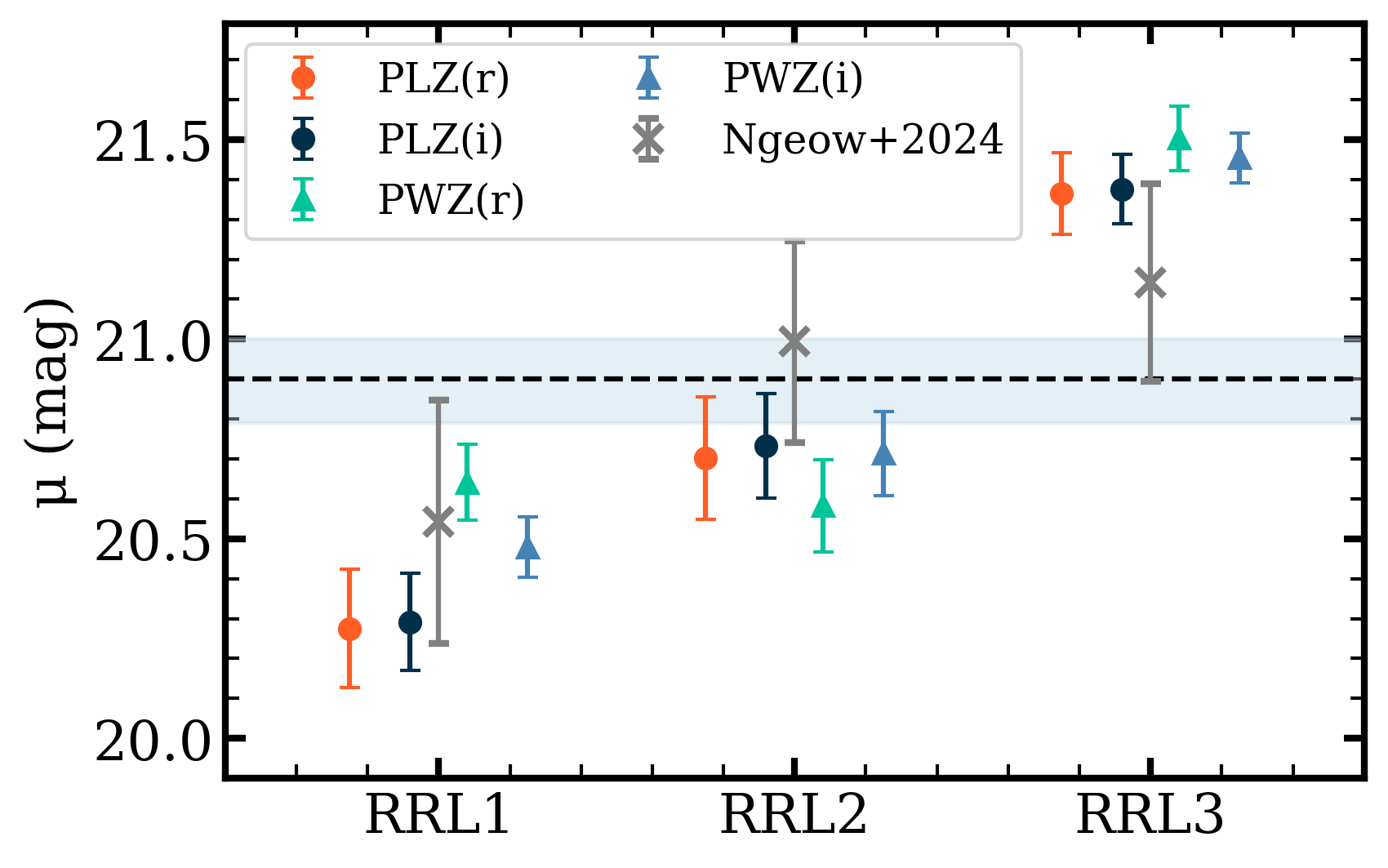}
    \caption{
    A comparison of the distance modulus estimates for each RR Lyrae star.
    The circles denote distance moduli estimated through the PLZ (Section~\ref{sec:rrl-plz}), the triangles show those estimated through the PWZ (Section~\ref{sec:rrl-pwz}) and the `$\times$' points show literature values.
    The PLZ and PWZ each contain two points for the $r$ and $i$ band values.
    The horizontal dashed line shows the distance modulus from the isochrone fit ($\mu = \satmod\,\si{mag}$).
    }
    \label{fig:rrl_distmods}
\end{figure}

Figure~\ref{fig:rrl_distmods} shows a comparison of all of the estimated distance moduli of each RR Lyrae.
All of our estimates are consistent with literature values from \citet{ngeow2024}.
Although we calculate distance moduli from PLZ and PWZ relations for RRL~3, the results for that particular star must be treated with caution, due to the high scatter in its light curve and the unphysical template fit.
We also note that while the RMS residual (column labeled `$\sigma$' in Table~4 of \citealt{marconi2022}) contributes to $\sim 20$--$50\%$ of the propagated uncertainty in the PWZ fits, it is the dominant component of the uncertainty in the PLZ fits, roughly doubling the estimated error.
\textbf{}

\section{Discussion}
\label{sec:discussion}

\begin{figure}[h!]
    \centering
    \includegraphics[width=0.95\columnwidth]{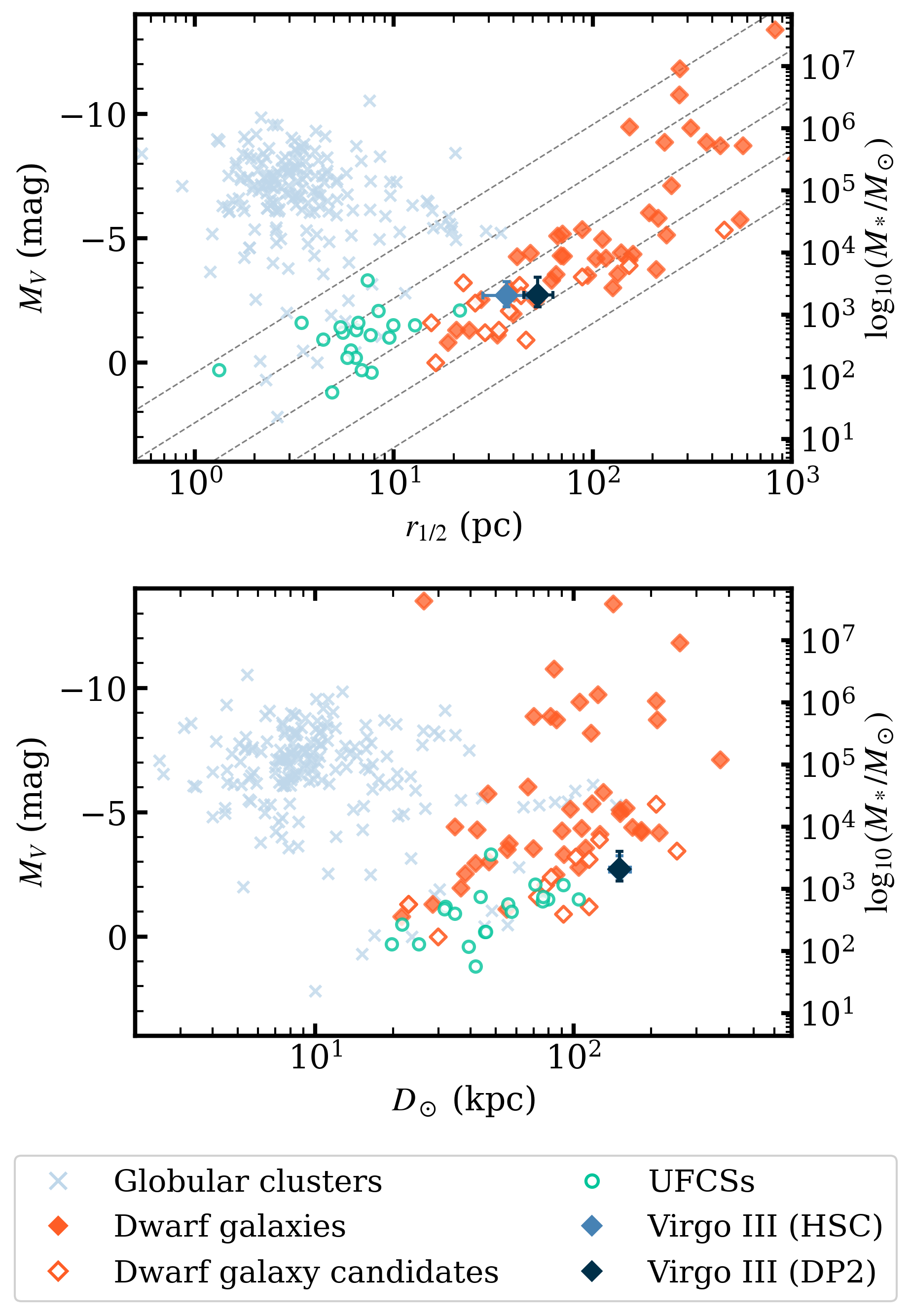}
    \caption{
    The top panel shows the absolute $V$-band magnitude of Milky Way satellite systems as a function of azimuthally averaged half-light radius. 
    The bottom panel shows the same, but as a function of heliocentric distance.
    The second $y$-axis on the right hand side of both panels shows stellar mass assuming $M_*/L_V = 2$.
    The light blue `$\times$' points show globular clusters, filled orange diamonds show confirmed dwarf galaxies, unfilled diamonds show candidate dwarf galaxies, and unfilled cyan circles show ambiguous ultra-faint compact satellites \citep[UFCSs;][]{cerny2026}.
    The Virgo~III measurement from DP2 is shown as a dark blue diamond with error bars, while the HSC measurement is shown as a lighter blue diamond.
    The dashed gray lines show contours of constant surface brightness (24, 26, 28, 30, 32 $\si{mag.arcsec^{-2}}$).
    }
    \label{fig:rhalf_dsun_Mv}
\end{figure}
Rubin DP2 represents the largest and deepest dataset released by the Rubin Observatory, covering $\sim$3,000\,deg$^2$ of the southern sky in six bands.
The deep Rubin DP2 data confirm that Virgo~III has structural parameters that are broadly consistent with the shallower HSC discovery measurements from \citet{homma2024}. 
The most notable improvement is greater than a factor of four increase in candidate member stars ($N_* = \satnstar$ vs.\ $25^{+5}_{-4}$).
This is a direct consequence of the greater depth of the DP2 coadded imaging, with $\sim$62.5\% of our high-purity member sample lying below the HSC magnitude limit of $i \lesssim 24.5$, allowing us to probe stars well below the main-sequence turnoff of Virgo~III.
We find that our measured heliocentric distance ($D_\odot = \satheliodist$) and $V$-band absolute magnitude ($M_V = \satmv$) are fully consistent with the HSC measurements.

The deeper DP2 data prefer a larger ellipticity ($\varepsilon = \satell$) and elliptical half-light radius ($a_h = \satah$), resulting in a larger azimuthally averaged physical half-light radius of $r_{1/2} = \satphysrh$.
 We attribute these discrepant measurements to the presence of a low-significance overdensity of faint stars extending southward from the core of Virgo~III.
This extended feature was likely unobservable in the HSC data because it is populated by stars that are fainter than the threshold applied by \citet{homma2024}.
The presence of this extended feature could be indicative of tidal disruption or an extended halo \citep[e.g.,][]{deason2014, chiti2021, pace2022}; however, further work is needed to confirm its reality.

Figure~\ref{fig:rhalf_dsun_Mv} shows the absolute $V$-band magnitude of currently known Milky Way satellite systems as a function of azimuthally averaged physical half-light radius and heliocentric distance.
We plot the previous measurements of Virgo~III with HSC (light blue diamond) and our new measurements from Rubin DP2 (dark blue diamond). The size and luminosity of Virgo~III place it among many candidate (unfilled orange diamonds) and spectroscopically confirmed (filled orange diamonds) UFDs.
In the absence of spectroscopic data, the morphology of Virgo~III (including the larger physical size measured here) suggests that it is likely a UFD.

The RGB candidate member stars of Virgo~III (that were cross-matched with \textit{Gaia}) show a prominent ($\sim$20\,mmag) offset from the more metal-rich foreground stellar locus in $g-r$ vs.\ $r-i$ color--color space. This provides qualitative evidence for a metal-poor population \citep{li2019}, which is consistent with the metallicity of $\rm [Fe/H] \lesssim \satfeh$ derived from isochrone fitting.
Spectroscopic follow-up of the brightest member stars would allow a direct metallicity and radial velocity measurement.
Deeper $u$-band photometry could also enable accurate photometric metallicity measurements for member stars.

The DP2 forced photometry light curves of the two RRab stars yield well-constrained periods and distances consistent with previously measured values, demonstrating the time-domain capabilities of Rubin.
The RRc star exhibits significant scatter and unphysical best-fit parameters, likely reflecting its shorter period and lower amplitude, which make it more susceptible to the $\sim$0.02--0.03\,mag systematic floor in the per-visit photometric calibration.

Overall, this analysis demonstrates that the Rubin DP2 data are of sufficient quality to detect and characterize faint stellar systems at $\sim$150\,kpc, and lie within expectations of the latest predictions mentioned in Section~\ref{sec:intro}. 
The methods developed here also provide a framework directly applicable to future LSST data releases.
These data, which were collected with only $\sim$10.5 hours of effective exposure time, approximate the depth that will be achieved over the 10-year LSST WFD program. We show that such depth is sufficient to recover the morphological, photometric, and time-domain properties of Virgo~III in unprecedented detail.
Similar approaches can be applied to characterize the many faint satellite systems that are expected to be discovered across the southern sky by Rubin LSST.

\section*{Acknowledgments}
AP acknowledges the support of the Eckhardt Fellowship at the University of and the Chateaubriand Fellowship.
This work is partially supported by the National Science Foundation under Grant No.\ AST-2006340, AST-2307126, and AST-2407526.
This research is partially funded by a
generous gift from Charles Simonyi to the NSF Division of Astronomical Sciences. The award is made in recognition of significant contributions to Rubin Observatory’s Legacy Survey of Space and Time.
WC gratefully acknowledges support from a Gruber Science Fellowship at Yale University.  This material is based upon work supported by the National Science Foundation Graduate Research Fellowship Program under Grant No.\ DGE2139841.

This material is based upon research supported by the Chateaubriand Fellowship of the Office for Science \&
Technology of the Embassy of France in the United States.

This material is based upon work supported in part by the National Science Foundation through Cooperative Agreements AST-1258333 and AST-2241526 and Cooperative Support Agreements AST-1202910 and 2211468 managed by the Association of Universities for Research in Astronomy (AURA), and the Department of Energy under Contract No. DE-AC02-76SF00515 with the SLAC National Accelerator Laboratory managed by Stanford University. Additional Rubin Observatory funding comes from private donations, grants to universities, and in-kind support from LSST-DA Institutional Members.

This research uses services or data provided by the Rubin Science Platform at NSF-DOE Vera C. Rubin Observatory, which is jointly funded by the U.S. National Science Foundation and the U.S. Department of Energy, Office of Science.

This work has made use of data from the European Space Agency (ESA) mission
{\it Gaia} (\url{https://www.cosmos.esa.int/gaia}), processed by the {\it Gaia}
Data Processing and Analysis Consortium (DPAC,
\url{https://www.cosmos.esa.int/web/gaia/dpac/consortium}). Funding for the DPAC
has been provided by national institutions, in particular the institutions
participating in the {\it Gaia} Multilateral Agreement.

This manuscript has been authored by Fermi Research Alliance, LLC under Contract No.\ DE-AC02-07CH11359 with the U.S. Department of Energy, Office of Science, Office of High Energy Physics. The United States Government retains and the publisher, by accepting the article for publication, acknowledges that the United States Government retains a nonexclusive, paid-up, irrevocable, world-wide license to publish or reproduce the published form of this manuscript, or allow others to do so, for United States Government purposes.

\facilities{Rubin:Simonyi, Rubin:USDAC, Gaia}
\software{Rubin Science Platform \citep{omullane2024}, astropy \citep{astropy2013, astropy2018, astropy2022}, jupyter \citep{jupyter2021}, matplotlib \citep{matplotlib2007}, numpy \citep{numpy2020}, pycycle, scipy \citep{scipy2020}, ugali \citep{bechtol2015,drlicawagner2020}.}

\bibliography{references}{}
\bibliographystyle{aasjournal}

\clearpage
\mbox{~}
\clearpage
\begin{appendix}

\section{MCMC Posterior Distributions}
\label{app:ugali_mcmc}
In this Appendix, we show the posterior distributions of a simultaneous fit of the morphological and isochrone parameters of Virgo~III using the \texttt{ugali} package, also described in Section~\ref{sec:morph}.
The posterior was sampled with the \texttt{emcee} package using 20 walkers with 15,000 steps after an initial burn-in of 1,000 steps.
\begin{figure*}[h!]
    \centering
    \includegraphics[width=0.95\textwidth]{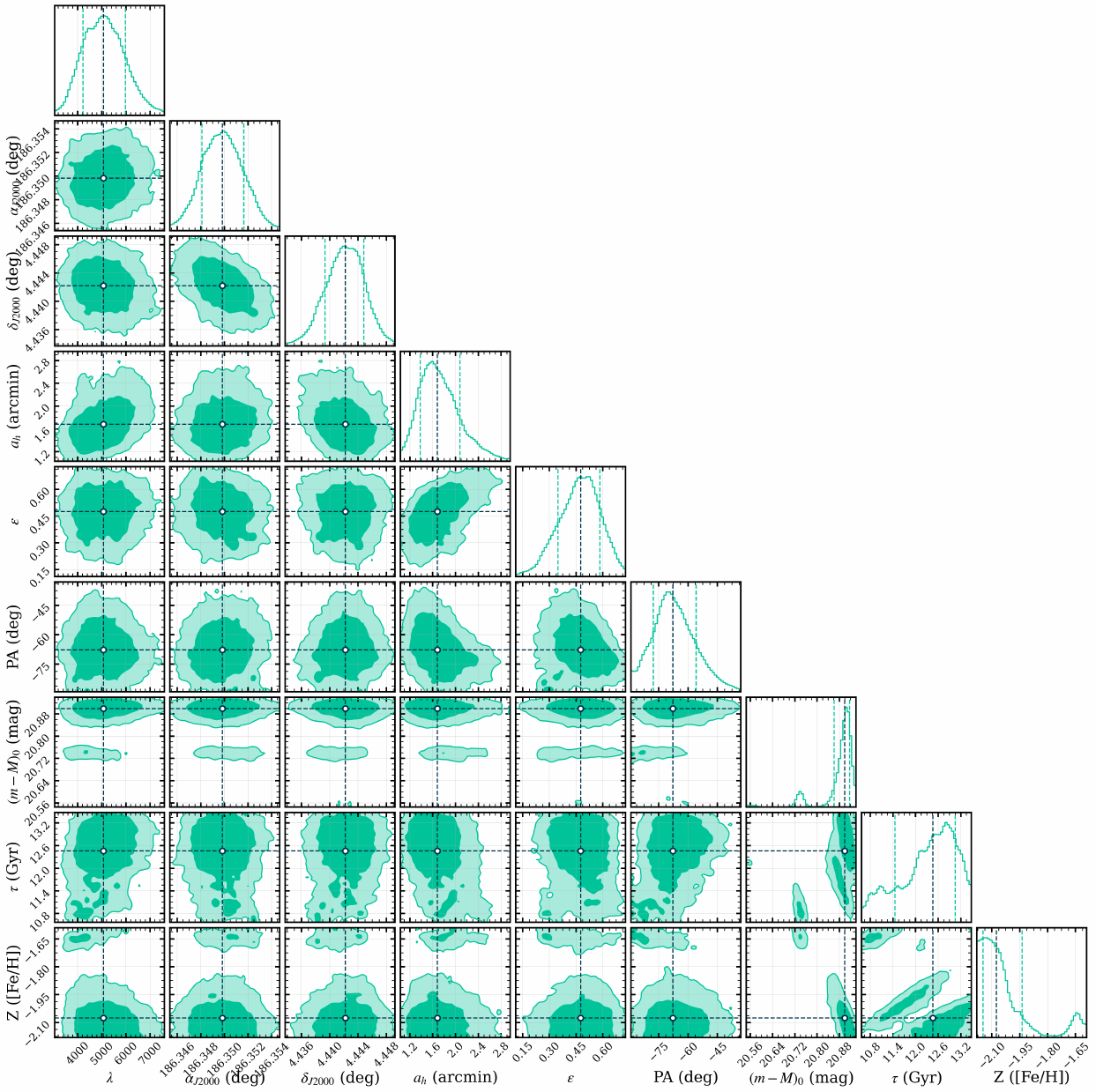}
    \caption{
    Posterior distributions of the morphological properties of Virgo~III modelled with \texttt{ugali}.
    Here, $\lambda$ is the stellar richness, $(\alpha_{J2000}, \delta_{J2000})$ are the centroid coordinates, $a_h$ is the angular semi-major axis of an ellipse containing half the light, $\varepsilon$ is the ellipticity, PA is the position angle, $(m-M)_0$ is the distance modulus, $\tau$ is the stellar age and $Z$ is the stellar metallicity.
    }
    \label{fig:mcmc}
\end{figure*}

\clearpage

\section{Other Isochrone Filters}
\label{app:isochrones} 
We show diagnostic plots of the matched-filter isochrone selection for Virgo~III, similar to Figure~\ref{fig:gr_6panel}. 
These figures use Rubin DP2 $g$-, $r$-, and $i$-band photometry and PARSEC-COLIBRI isochrones generated in the LSST bandpasses.
\begin{figure*}[h!]
    \centering
    \includegraphics[width=0.90\textwidth]{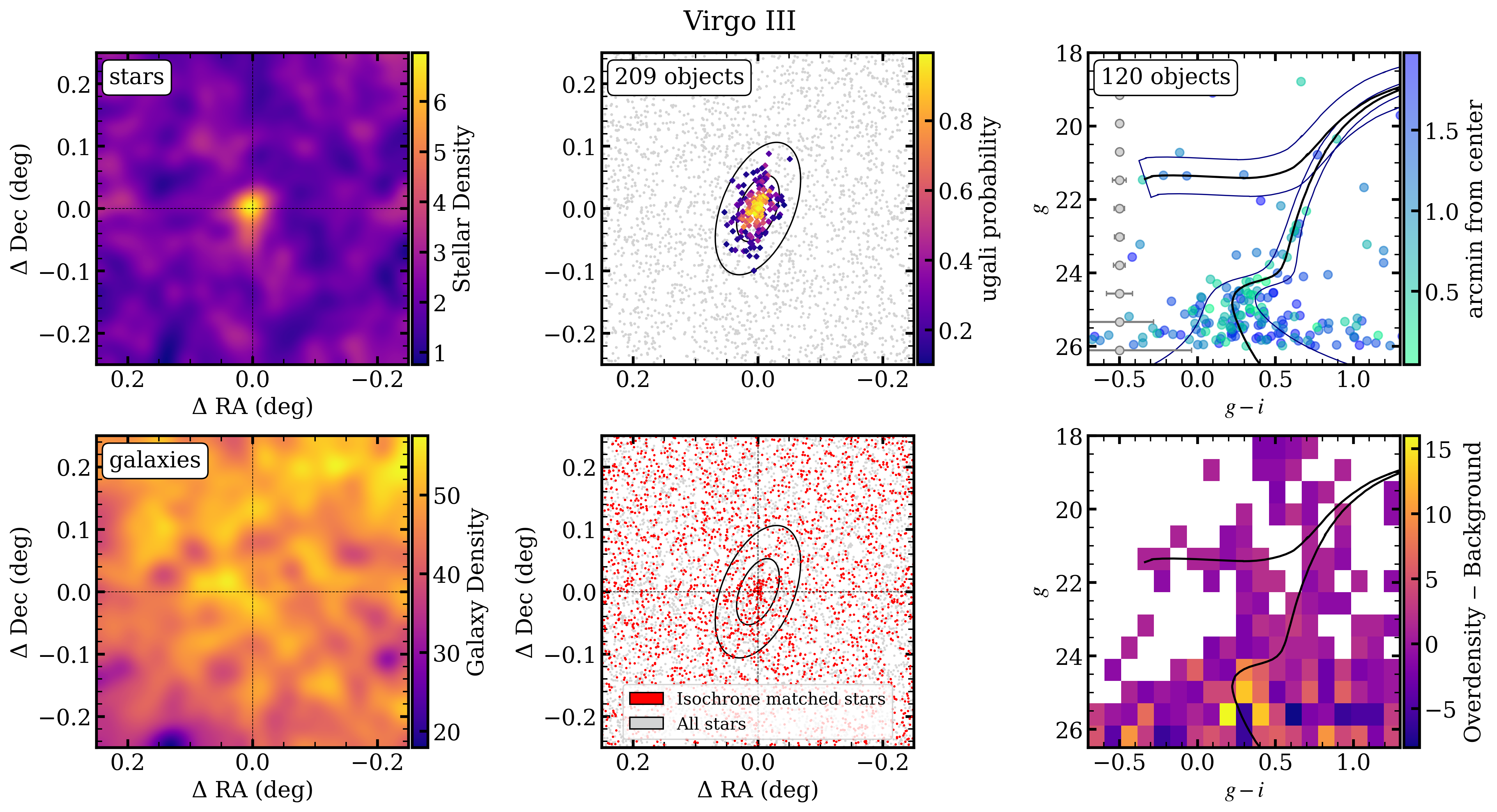}
    \caption{Diagnostic plots for Virgo~III similar to Figure~\ref{fig:gr_6panel}, but with a $g$ vs.\ $g-i$ isochrone filter.}
    \label{fig:gi_6panel}
\includegraphics[width=0.90\textwidth]{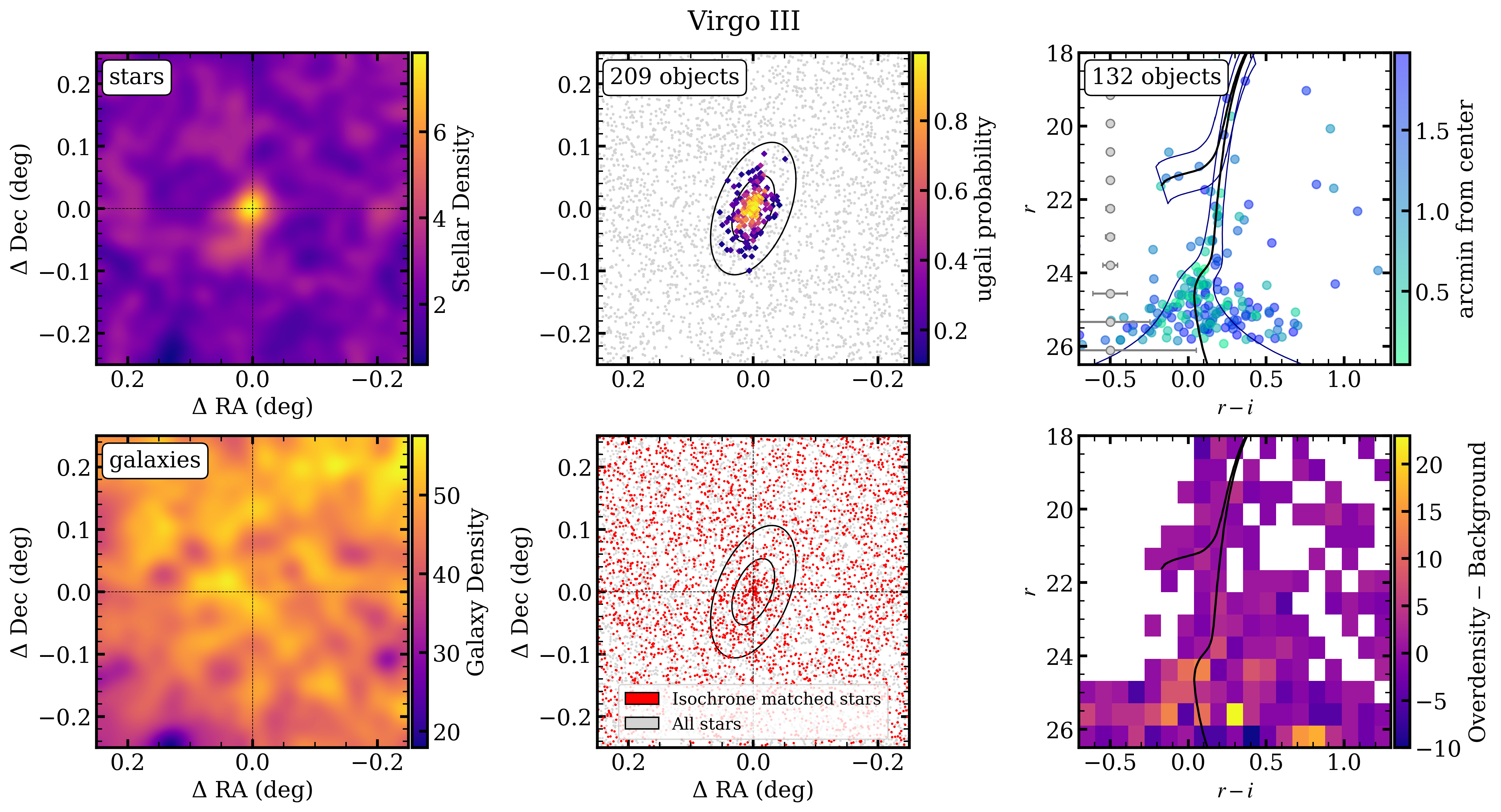}
    \caption{Diagnostic plots for Virgo~III similar to Figure~\ref{fig:gr_6panel}, but with a $r$ vs.\ $r-i$ isochrone filter. 
    Note that the $x$-axis limits of the CMDs in the rightmost panels are different from those in Figures~\ref{fig:gr_6panel} and \ref{fig:gi_6panel}.}
    \label{fig:ri_6panel}
\end{figure*}
\end{appendix}
\end{document}